\documentclass[fleqn,usenatbib]{mnras}

\usepackage{etoolbox}

\usepackage{newtxtext,newtxmath}

\usepackage[T1]{fontenc}

\DeclareRobustCommand{\VAN}[3]{#2}
\let\VANthebibliography\thebibliography
\def\thebibliography{\DeclareRobustCommand{\VAN}[3]{##3}\VANthebibliography}

\usepackage{graphicx}	
\usepackage{amsmath}	
\usepackage{placeins}
\usepackage{float}
\usepackage{natbib}
\usepackage{xcolor}

\defcitealias{Yoon2019}{Y19}
\defcitealias{Yang2007}{Y07}

\newcommand{\ar}{\alpha_{R}}
\newcommand{\afiv}{\alpha_{5}}

\newcommand{\degt}{\delta_{8}}

\newcommand{\BT}{\mathrm{B}\slash\mathrm{T}}
\newcommand{\hmpc}{h^{-1}\mathrm{Mpc}}
\newcommand{\hmsol}{h^{-1}M_\odot}
\newcommand{\hhmsol}{h^{-2}M_\odot}

\newcommand{\sma}{\mathrm{SMA}}
\newcommand{\kpc}{\mathrm{kpc}}
\newcommand{\eb}{e_\mathrm{bar}}
\newcommand{\epk}{e_\mathrm{peak}}
\newcommand{\rpk}{\mathrm{SMA}_\mathrm{peak}}
\newcommand{\pb}{\mathrm{P_{GZ2}}}
\newcommand{\meb}{\langle e_\mathrm{bar} \rangle}
\newcommand{\delmeb}{\Delta \langle e_\mathrm{bar} \rangle}

\title[Tidal Dependence of Galactic Bars]{Dependence of galactic bars on
the tidal density field in the SDSS}

\author[Deng et al.]{
Qi'an Deng,$^{1}$
Ying Zu,$^{1,2,3}$\thanks{E-mail: yingzu@sjtu.edu.cn}
Shadab Alam,$^{4}$
and
Yongmin Yoon$^{5}$
\\
$^{1}$Department of Astronomy, School of Physics and Astronomy, Shanghai Jiao Tong
University, Shanghai 200240, China\\
$^{2}$Shanghai Key Laboratory for Particle Physics and Cosmology, Shanghai Jiao Tong
University, Shanghai 200240, China\\
$^{3}$Key Laboratory for Particle Physics, Astrophysics and Cosmology,
Ministry of Education, Shanghai Jiao Tong University, Shanghai 200240,
China\\
$^{4}$Tata Institue of Fundamental Research, Homi Bhabha Road, Mumbai
400005, India\\
$^{5}$Korea Astronomy and Space Science Institute (KASI), 776 Daedeokdae-ro, Yuseong-gu, Daejeon, 34055, Republic of Korea
}

\date{Accepted XXX. Received YYY; in original form ZZZ}

\pubyear{2023}

\begin{document}
\label{firstpage}
\pagerange{\pageref{firstpage}--\pageref{lastpage}}
\maketitle

\begin{abstract}
    As a key driver of the secular evolution of disc galaxies, bar
    formation is potentially linked to the surrounding tidal field.  We
    systematically investigate the dependence of bars on both the
    small~(${<}2\,\mathrm{Mpc}/h$) and large~(${>}5\,\mathrm{Mpc}/h$) scale
    tidal fields using galaxies observed between $0.01{<}z{<}0.11$ by the
    Sloan Digital Sky Survey~(SDSS). We characterise bar strength using the
    ellipticity of the isophote that corresponds to each bar,
    $e_{\mathrm{bar}}$, derived from its galaxy image after subtracting the
    2D disc component. We demonstrate the efficacy of our bar detection
    method by performing an extensive comparison with the visual
    identifications from SDSS and the DESI Legacy Surveys. Using the Yang
    et al. SDSS group catalogue, we confirm the results from a recent study
    that the average $e_{\mathrm{bar}}$ of galaxies within interacting
    clusters is higher than that within isolated ones at $0.01{<}z{<}0.06$,
    but this small-scale tidal enhancement of bars disappears after we
    increase the cluster sample by a factor of five to $z{=}0.11$. On large
    scales, we explore the dependence of $e_{\mathrm{bar}}$ on
    $\alpha_{5}$, the tidal anisotropy of the density field defined over
    $5\,\mathrm{Mpc}/h$.  We do not detect any such dependence for $98\%$ of the
    galaxies with $\alpha_{5}{<}10$. Intriguingly, among the $2\%$ with
    $\alpha_{5}{\ge}10$, we detect some hint of a boost in bar strength in
    the underdense regions and a suppression in the overdense regions.
    Combining our results on both scales, we conclude that there is little
    evidence for the tidal dependence of bar formation in the local Universe,
    except for the extremely anisotropic environments.
\end{abstract}

\begin{keywords}
methods: statistical --- galaxies: bar --- galaxies: disc --- galaxies: haloes --- galaxies: structure --- large-scale structure of the Universe
\end{keywords}



\section{Introduction}
\label{sec:intro}

As the most common non-axisymmetric structure in spiral galaxies, bars
serve as a potent agent for redistributing angular momentum in the
disc-halo systems, thereby driving the secular evolution of disc
galaxies~\citep{Kormendy2004,Sellwood2014}. {\it N}-body simulations have
shown that the formation of bars is almost an inevitable consequence of the
strong $m{=}2$ instability in disc galaxies, and once formed, bars are
believed to be long-lived~\citep[but see][]{Bournaud2002}. Yet, one third of
the discs in the local Universe remain
barless~\citep{deVaucouleurs1963,Barazza2008,Aguerri2009,Nair2010,Masters2011}.
One possibility is that the promotion and/or inhibition of bar-formation
instability may depend sensitively on the tidal density environment, which
is intimately linked to the halo angular momentum in the
$\Lambda$-dominated cold dark matter~($\Lambda$CDM) Universe.  In this
paper, we develop an automated bar detection method to measure the bar
strength for galaxies observed by the Sloan Digital Sky
Survey~\citep[SDSS;][]{York2000}. By examining the tidal field on both
small~(${<}2\,\hmpc$) and large~(${>}5\,\hmpc$) scales, we aim to
systematically search for the tidal dependence of bars in the local
Universe.

Despite the lack of a complete theory of bar formation~
\citep[see][for a recent review]{Sellwood2014}, the basic mechanism is
reasonably well understood~\citep{Binney2008}.  In the linear regime,
self-gravitating cold discs are globally unstable against non-axisymmetric
modes~\citep{Kalnajs1972}, resulting in the formation of weak spiral
disturbances. In a differentially rotating disc, these initial disturbances
grow by successive swing amplifications at the co-rotation radius and
reflections off the galactic centre~\citep{Toomre1981}. After a nascent bar
emerges out of this feedback loop, it could trap the eccentric stellar
orbits and force them to precess coherently, thereby reinforcing the
stellar bar~\citep{Lynden-Bell1979,Earn1996}. The bar-formation instability
generally sets off
spontaneously~\citep[][]{Hohl1971,Ostriker1973,Lynden-Bell1979,Sellwood1981,Efstathiou1982},
but can be triggered by an external perturber in mergers or flybys as
well~\citep[][]{Byrd1986,Noguchi1987,Noguchi1988,Gerin1990,Miwa1998,Berentzen2004,Martinez-Valpuesta2017,Peschken2019,Cavanagh2022}.
Numerical studies suggested that the presence of a hot
``spheroidal''~(a massive central bulge or dark matter halo) component may
stabilize the disc against bar
instability~\citep{Ostriker1973,Efstathiou1982,Kataria2018,Kataria2020,Jang2023}. However, a strong bar can
still form in the disc embedded in a massive ``live'' halo that responds to
the stellar
components~\citep{Hernquist1992,Debattista2000,Athanassoula2002,Athanassoula2003}.

Beyond halo mass, the spin of the dark matter haloes may also affect the
formation and growth of bars. By modelling the dynamical friction of a
rotating rigid bar inside an isothermal halo, \citet{Weinberg1985}
discovered that the halo spin can promote bar growth as the excess torque
of the prograde orbits helps increase the rate at which the bar loses its
angular momentum. More recently, \citet{Saha2013} carried out a suite of
{\it N}-body simulations of barred galaxies living in halos with different
amount of spin.  Confirming the semi-analytic arguments from
\citet{Weinberg1985}, they found that the bars form more rapidly and grow
stronger in co-rotating haloes with larger spin. During the subsequent
secular evolution, bar growth may instead be suppressed in fast-spinning
haloes that are capable of replenishing angular momentum to the
bar~\citep[][]{Long2014,Collier2018}.  Some cosmological hydrodynamic
simulations also predict a relatively lower frequency of barred galaxies in
fast-spinning haloes~\citep[][]{Rosas-Guevara2022,Izquierdo-Villalba2022},
despite their difficulty in resolving the short bars~\citep{Zhao2020}.  Yet
a high spin of the inner halo could trigger an earlier buckling, producing
a more pronounced boxy/peanut shape of the final bar~\citep{Kataria2022}.
Therefore, despite the lack of consensus on the sign of the effect,
theories and simulations both predict that the bar strength of disc
galaxies depends on the halo spin.

While inaccessible observationally\footnote{It is plausible to measure the
angular momentum distribution of gas inside clusters using future
Sunyaev-Zeldovich surveys~\citep{Cooray2002, Baxter2019}.}, halo spin can
be indirectly probed using the tidal field inferred from the 3D
distribution of galaxies, assuming a tidal origin of the halo angular
momentum. During the structure formation under $\Lambda$CDM, it is widely
accepted that each halo acquired its angular momentum during the linear
growth stage from the tidal torques induced by the surrounding matter
distribution, as predicted by the tidal torque
theory~\citep{Peebles1969,Doroshkevich1970,White1984}.  This linear channel
of spin growth eventually shut down when the overdensity collapsed into a
halo~\citep{Porciani2002}. During the subsequent nonlinear evolution, the
overall spin growth is sustained by the coherent accretion and mergers,
during which the orbital angular momentum of the approaching systems is
transferred into the spin of the
halo~\citep{Gardner2001,Maller2002,Vitvitska2002}.  This nonlinear channel
of spin growth is most likely induced by the tidal torques on small
scales~\citep{Hetznecker2006}.

Therefore, both the linear and non-linear channels rely on the strong tidal
field to spin up dark matter haloes, but with the field defined on
different scales. Close pairs of galaxy clusters are among the most
plausible sites for enhanced tidal torque on small scales.  Recently,
\citet[][hereafter referred to as~\citetalias{Yoon2019}]{Yoon2019} measured
the barred galaxy fraction in $105$ clusters between redshift $0.015$ and
$0.06$, finding a significant enhancement of bar fraction among galaxies
surrounding the interacting clusters over the isolated ones. This is a
tentative observational evidence that the bar growth is boosted by the
small-scale tidal torque.  However, the cluster sample used by
\citetalias{Yoon2019} is relatively small. In this work, we elucidate the
small-scale tidal dependence of bars by performing a similar analysis over
a significantly larger sample of clusters with redshifts up to $0.11$.

Meanwhile, the correlation between the spin of haloes and their large-scale
tidal field has been robustly detected in cosmological simulations. For
instance, using a suite of high-resolution {\it N}-body
simulations~\citep{Jing2007}, \citet{Wang2011} computed a force-based tidal
field for each halo by summing up the tidal forces exerted by all other
haloes above $10^{12}\,\hmsol$. They found that the haloes have a tendency to
spin faster in a stronger tidal field and the trend is stronger for more
massive haloes. Circumventing the need to compute forces,
\citet{Paranjape2018} proposed a tensor-based tidal anisotropy parameter
$\ar$ to quantify the strength of the tidal environment.
\citet{Ramakrishnan2019} later showed that $\ar$ correlates strongly with
halo spin, making it an excellent proxy of halo spin.  \citet{Alam2019}
subsequently demonstrated that $\ar$ can be robustly measured from the
observed number density distribution of galaxies.  While the dependence of
bar strength on the overdensity environment has been measured~\citep[albeit
with inconclusive results;][]{Aguerri2009, Lee2012, Li2009, Skibba2012,
Fraser-McKelvie2020}, few studies focused on the tidal density field.  In
this work, we use $\ar$ as a proxy for halo spin and examine the dependence
of bar strength on the tidal anisotropy of the galaxy density field {\it at
fixed overdensity}, aiming to detect the potential effect of large-scale
tidal field on the formation of bars.

A prerequisite to robustly detecting the tidal dependence of bars is an
accurate method of identifying bars and quantifying the bar strength.
There are mainly four types of bar identification methods in observations,
including visual inspection
\citep[e.g.,][]{deVaucouleurs1963,Nair2010,Masters2011}, ellipse
fitting~\citep{Wozniak1995,Jogee2004,Laine2002,Marinova2007,Aguerri2009,Menendez-Delmestre2007,Li2011},
Fourier
analysis~\citep{Elmegreen1985,Athanassoula2002,Athanassoula2003,Aguerri2009},
and 2D image
decomposition~\citep{Prieto2001,Aguerri2005,Laurikainen2005,Gadotti2008}.
In particular, visual inspection by citizen scientists through the Galaxy
Zoo project has been tremendously successful in robustly identifying barred
galaxies in the local Universe. At high redshifts, however, the reliability
of visual inspection deteriorates due to the reduction in the image quality
and spatial resolution. In this work, we apply the ellipse fitting method
to identify bars and measure their strength for a large sample of disc
galaxies up to redshift $z{=}0.11$. \citet{Gadotti2008} pointed out that
the bar ellipticity measured from ellipse fitting can be underestimated if
the disc component is relatively luminous. To overcome this, we develop an
automated bar detection method that performs ellipse fitting over galaxy
images after subtracting the 2D disc components, thereby significantly
improving our bar detection capability at the high redshifts.

The paper is organised as follows. we describe our various datasets in
Section~\ref{sec:data} and the measurement of tidal density environments in
Section~\ref{sec:tidal}. In Section~\ref{sec:det}, we present our automated
bar detection method based on ellipse fitting over the disc-subtracted
images of galaxies. The main results on the tidal dependence of bars are
presented in Section~\ref{sec:tidalbias}. We conclude by summarising our
results and looking to the future in Section~\ref{sec:conc}. We assume a
flat $\Lambda\mathrm{CDM}$ cosmology with $\Omega_\mathrm{m}=0.27$ and
$h=0.7$ for distances. Throughout this paper, we use $\hmsol$ and $\hhmsol$
as the units of the halo and stellar mass, respectively.

\begin{figure*}
	\includegraphics[width=\textwidth]{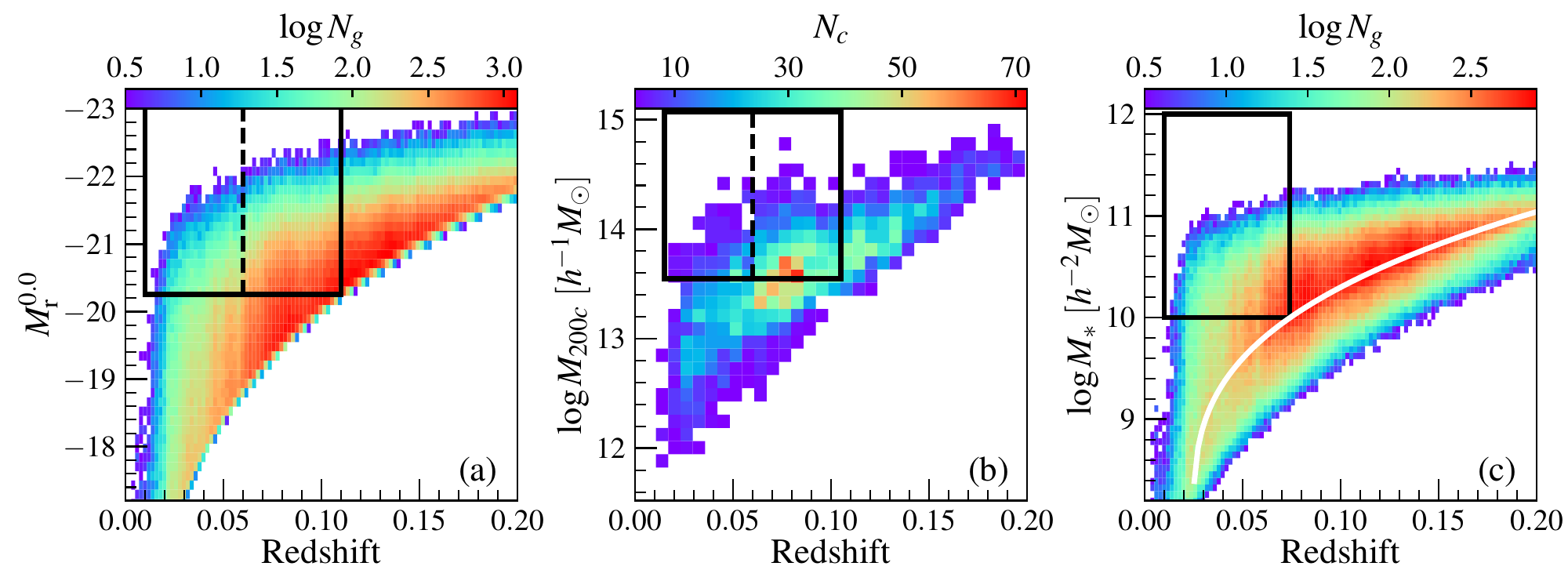}
    \caption{Galaxy and cluster samples selected for our analysis in the
    paper. {\it Panel (a)}: Number distribution of SDSS main sample
    galaxies on the r-band absolute magnitude~({\it K}-corrected to $z{=}0.0$)
    vs. redshift plane, colourcoded by the colourbar on top.  Thick black
    box indicates the luminosity-limited galaxy sample used by our
    small-scale tidal analysis, split by the vertical dashed line at
    $z{=}0.06$ into the low- and high-redshift subsamples.  {\it Panel
    (b)}: Similar to Panel (a), but for the \citetalias{Yang2007}
    groups/clusters on the halo mass vs. redshift plane. {\it Panel (c)}:
    Similar to Panel (a), but for the stellar mass-limited galaxy sample
    used by our large-scale tidal analysis.
	White solid curve indicates the
    stellar mass threshold above which the SDSS observation is roughly
    complete.}
	\label{fig:sample}
\end{figure*}

\section{Data}
\label{sec:data}

Our analyses in this work are primarily based on the main galaxy
sample~\citep{Strauss2002} of the Sloan Digital Sky Survey Data Release
7~\citep{Abazajian2009}.  For the parent galaxy sample for bar detection,
we construct a luminosity-limited volume-complete sample from the
\texttt{bbright0} sample in the NYU-VAGC catalogue~\citep[][]{Blanton2005},
including $131455$ galaxies with redshift $0.01{\leq} z{\leq} 0.11$ and r-band
absolute magnitude brighter than
$M_\mathrm{r}^{\mathrm{lim}}{=}{-}20.25$~(K-corrected to $z{=}0$ following
\citet{Blanton2007}). The values of the maximum redshift $z_{\mathrm{max}}$
and magnitude limit are chosen so that a typical bar in the galaxy with
$M_\mathrm{r}{\simeq}{-}20.25$, which is around $3\,\kpc$ in length, can be
roughly resolved by the SDSS r-band imaging at $z_{\mathrm{max}}$ with a
median seeing of 1.32 arc-seconds, which corresponds to $2.7\,\kpc$ at
$z{=}0.11$. In addition, the bar size of galaxies brighter than ${-}20.25$
increases rapidly with stellar mass as $M_*^{0.6}$~\citep{Erwin2019},
rendering most of bars detectable in our sample.  From the parent
luminosity-limited sample, we remove $31010$ highly-inclined discs that are
unfit for bar detection by requiring their outer ellipticity
$e_{90}{<}0.5$, where $e_{90}$ is defined as the ellipticity of the
isophote which encloses
$90\%$ of the total luminosity. We measure $e_{90}$ using the isophote
fitting described in \S\ref{subsec:disc_sub}.
In total, we have $100445$ galaxies that are roughly face-on in the
luminosity-limited sample.

When measuring the bar strength in \S\ref{sec:det}, we minimize the
contamination from neighbouring galaxies~(e.g., when subtracting the 2D
disc) by using the r-band SDSS atlas~\citep{Stoughton2002} images
downloaded from the SDSS Data Archive Server~(DAS). Each atlas image stamp
only includes the light originated from the galaxy in the centre of the
stamp, therefore providing a clean 2D surface brightness~(SB) distribution
for ellipse fitting.  In order to perform the disc subtraction~(see
\S\ref{subsec:disc_sub} for details), we adopt the structural parameters
measured by \citet{Simard2011} from the r-band image of each galaxy,
including the bulge-to-total ratio~($\BT$), the effective radius and
position angle~(PA) of the bulge, the scale length and PA of the disc, and
the inclination angle.  The fractional uncertainties in the measurements of
the disc inclination and
PA are both around $2{-}3\%$, producing robust 2D disc models for the
majority of galaxies in our sample.
In addition, for the visual validation of our bar
detection method, we also present the galaxy images from the DESI DR9
imaging~\citep{Dey2019}, which are ${\sim}1.4$ magnitudes deeper than SDSS
in the {\it r}-band. We plan to directly apply our bar detection method to the
DESI imaging data in a follow-up paper~(Deng et al. {\it in prep}). We
construct the color composite SDSS and DESI images for each galaxy
using the method described in \citet{Lupton2004}.

\begin{figure*}
	\includegraphics[width=\textwidth]{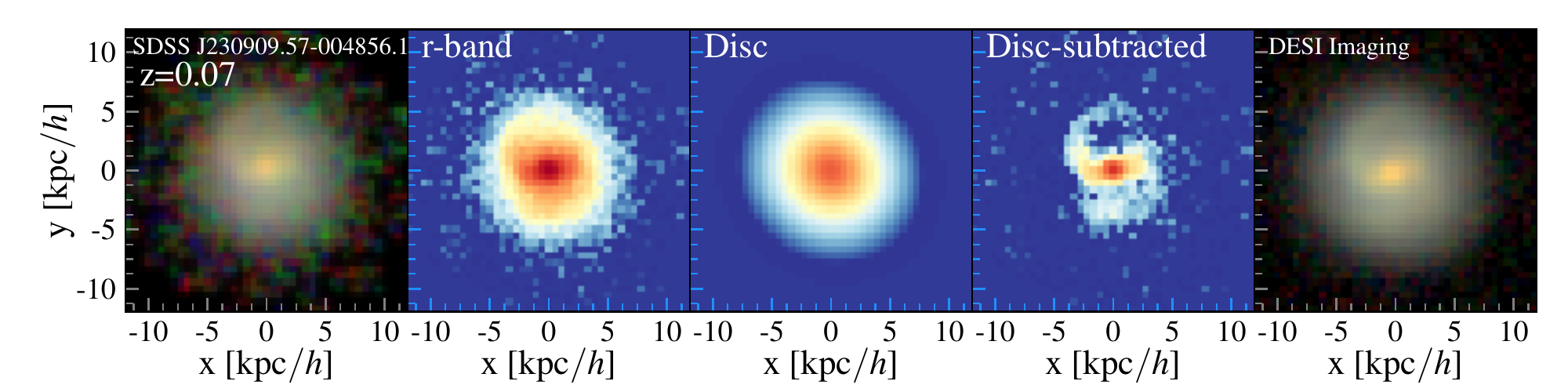}
	\caption{Five different types of images of an example galaxy at
	$z{=}0.07$,
    including~(from left to right) the SDSS {\it gri} colour
	composite image, original SDSS {\it r}-band image, best-fitting 2D
	disc model, disc-subtracted SDSS {\it r}-band image, and the DESI
	{\it grz} colour composite image.
    The bar-like structure is reasonably recognisable in the DESI image,
    but is not revealed in the SDSS images until after the disc
    subtraction. }
	\label{fig:disc_subtraction_example}
\end{figure*}

For the comparison with visual inspection, we make use of the results from
the second phase of the Galaxy Zoo~(GZ2) project~\citep{Willett2013}. As
one of the most popular citizen science projects in Astronomy, GZ2 asks
volunteers from the public to classify the morphology of SDSS galaxies
through a carefully constructed decision tree. In essence, the volunteers
were asked to choose an answer to a descriptive question on the galaxy
morphology at each step along the decision tree. The fraction of people
choosing a particular answer to one question for the galaxy can be
interpreted as the probability for the galaxy to have that particular
morphological feature. The large number of galaxies with probabilistic
visual classifications makes GZ2 one of the most valuable galaxy morphology
catalogue to date. For our comparison, we use the
consistency-weighted~\citep[see section 3.2 of][]{Willett2013} bar fraction
as the probability of a galaxy hosting a bar in GZ2 and denote it as $\pb$.
We are aware of the latest GZ2 product using the DESI DECaLS
images~\citep{Walmsley2022}, which however includes galaxies in a smaller
footprint than SDSS and is thus less useful for our comparison.

To identify strong tidal environments on small scales, we select
interacting pairs of clusters from the spectroscopic group catalogue of
\citet[][hereafter referred to as~\citetalias{Yang2007}]{Yang2007}. The
\citetalias{Yang2007} groups were identified by applying an adaptive
halo-based group finder to the SDSS main galaxy sample, and we refer
interested readers to \citet{Yang2005,Yang2007} for the technical details.
We assume the brightest cluster galaxies~(BCGs) as the centres of the
clusters, as weak lensing studies showed that the BCGs are generally better
centroids than the default luminosity-weighted centres~\citep{Wang2022,
Golden-Marx2022}. \citetalias{Yang2007} provided a halo mass estimate
$M_{180m}$ for each group using abundance matching. To facilitate our
comparison with the \citetalias{Yoon2019} results, we convert $M_{180m}$
into $M_{200c}$ using the fitting formula from \citet{Hu2003}. We also make
sure each selected cluster has at least $5$ spectroscopic member galaxies
found by the \citetalias{Yang2007} catalogue. After selection, our sample
includes $1501$~($600$) clusters with $\log M_{200c}{\ge} 13.55$~($13.85$)
between redshift $0.015$ and $0.105$.  The redshift range of the cluster
sample is slightly narrower than that of the luminosity-limited galaxy
sample, thereby providing a complete coverage of the galaxy environment
surrounding the clusters at the bin edges. Our selection results in $109$
clusters with mass above $\log M_{200c}{=}13.85$ in the redshift range of
$0.015{<}z{<}0.06$, similar to the sample size in the \citetalias{Yoon2019}
analysis that included $105$ clusters within the same redshift range.
However, by extending the analysis up to $0.105$, we probe a comoving
volume that is about six times that of the \citetalias{Yoon2019} analysis,
yielding $600$ clusters above $\log M_{200c}{=}13.85$ within our full
redshift range of $0.015{<}z{<}0.105$.

For measuring the large-scale tidal anisotropy parameter, we follow the
method described by \citet{Alam2019} and select a stellar mass-limited
sample with $z \in [0.01,0.074]$ and $\log M_{*}{>}10$ as the tracer of the
underlying dark matter density field. For the seven per cent galaxies
without redshift due to fibre collision, we assign them the redshifts of
their nearest neighbours.  To minimize the boundary effect in calculating
the tidal field, we only include galaxies within the contiguous area in the
North Galactic Cap and those in regions with angular completeness larger
than $0.8$. We adopt the stellar mass estimates from the latest MPA-JHU
value-added galaxy catalogue, derived following the philosophy of
\citet{Kauffmann2003} and \citet{Brinchmann2004} assuming the
Chabrier~\citep{Chabrier2003} initial mass function and the
\citet{Bruzual2003} stellar population synthesis model. Based on the
stellar mass completeness limit estimated in \citet{Zu2015}, we expect the
SDSS main galaxy sample to be roughly volume-complete within $z{=}0.074$ down
to $\log M_{*}{=}10$. We choose a lower maximum redshift~($0.074$) than
that of the luminosity-limited one~($0.11$), because the tidal anisotropy
calculation requires a higher galaxy number density.  In total, the stellar
mass-limited sample includes $65222$ galaxies.

Figure~\ref{fig:sample} summarises the samples used by the analyses in the
current work, including the luminosity-limited galaxy sample,
\citetalias{Yang2007} cluster sample, and the stellar mass-limited galaxy
sample, inside the thick black box shown in the left, middle, and right
panel, respectively. In each panel, the 2D histogram indicates the number
distribution of all SDSS galaxies~(clusters) on the redshift vs.
stellar~(halo) mass plane, colour-coded by the colourbar on top.  For the
left and middle panels, each black dashed vertical line indicates the
redshift~($0.06$) that we use to split the overall sample into the low- and
high-redshift subsamples. The white curve in the right panel indicates the
mixture limit derived by \citet{Zu2015}, above which the galaxy sample
should be roughly complete in stellar mass.

\section{Measurement of Tidal Density Environments}
\label{sec:tidal}

\subsection{Small Scales: Interacting Cluster Pairs}
\label{subsec:clustersplit}

Following \citetalias{Yoon2019}, we classify each cluster pair as
interacting vs. non-interacting based on their projected physical
separation and the line-of-sight velocity difference. In particular, an
interacting pair of clusters should have the projected separation $D$
smaller than twice the sum of the two cluster radii $R_{200c}$, and the
velocity difference $\Delta v$ smaller than $750\,\mathrm{km}/s$. After
applying those criteria, we identify $274$ clusters in interacting pairs
among the $1501$ clusters in our overall sample with $\log M_{200c}{\ge}
13.55$, and the numbers reduce to $138$ interacting clusters out of the
$600$ clusters above $\log M_{200c}{=}13.85$. Splitting the cluster sample
above $\log M_{200c}{=}13.55$~($13.85$) by redshift, we find $47$~($25$)
interacting clusters among the $309$~($109$) clusters within
$z\,{=}[0.015,0.06]$ and $227$~($113$) interacting out of $1192$~($491$)
within $z\,{=}[0.06,0.105]$. Within the low-redshift bin that overlaps with
the \citetalias{Yoon2019} sample, we recover $91$ out of their $105$
clusters and all $16$ of their clusters in interacting pairs.

The membership criteria adopted by the \citetalias{Yang2007} group finder
are relatively stringent. We instead follow the practice of
\citetalias{Yoon2019} and re-define the member galaxies associated with
each cluster as those projected within $R_{200c}$ from the BCG and have
line-of-sight velocities relative to the BCG within $\pm 3\sigma_{v}$,
where $\sigma_v$ is the line-of-sight velocity dispersion of the cluster so
that $\sigma^{2}_{v}=G M_{200c}/(3R_{200c})$. We will adopt this new
membership definition for the small-scale tidal analysis in
\S\ref{subsec:cluster_scale}.

\subsection{Large Scales: Tidal Anisotropy Parameter}
\label{subsec:tidalenv}

We quantify the strength of the large-scale tidal field using the tidal
anisotropy parameter $\alpha_R$, first introduced by \citet{Paranjape2018}
as a measure of the anisotropic level of the tidal field. The original
definition of $\alpha_R$ is
\begin{equation}
    \alpha_R=\sqrt{q_R^2}(1+\delta_R)^{-1},
    \label{eqn:alphar}
\end{equation}
where $\delta_R$ is the spherical overdensity within a
sphere of radius {\it r} centred on the halo/galaxy and $q_R^2$ is the tidal shear
that can be computed as~\citep{Heavens1988,Catelan1996}
\begin{equation}
    q_R^2=\frac{1}{2}[(\mathrm{\lambda_3}-\mathrm{\lambda_2})^2+(\mathrm{\lambda_2}-\mathrm{\lambda_1})^2+(\mathrm{\lambda_3}-\mathrm{\lambda_1})^2],
\end{equation}
where $\lambda_1{<}\lambda_2{<}\lambda_3$ are the eigenvalues of the tidal
tensor. Following \citet{Alam2019}, we use a variant of
Equation~\ref{eqn:alphar}~(as will be shown further below) to measure the
tidal anisotropy parameter over the galaxy density field smoothed over
$5\,\hmpc$~(hereafter referred to as $\afiv$). We briefly summarize the
procedure for calculating $\afiv$ and refer interested readers to
\citet{Alam2019} for the technical details. Firstly, we calculate the
density field $\delta$ at each galaxy's position as the inverse of the
volume associated with that galaxy derived using the Voronoi tessellation
technique. We then interpolate the density field on a regular Cartesian
grid and smooth the density field using a Gaussian kernel of width
$5\,\hmpc$ to remove discontinuities near the masked regions or survey
boundaries; Secondly, we derive the gravitational potential from the
overdensity field by solving the Poisson equation and then compute $q_5$
using the components of the tidal tensor derived in the Fourier space;
Lastly, we calculate $\afiv$ via
\begin{equation}
    \alpha_5=\sqrt{q_5^2}(1+\delta_5)^{-0.55},
    \label{eqn:afiv}
\end{equation}
where we have changed the power-law index from $-1$ in
Equation~\ref{eqn:alphar} to $-0.55$. As demonstrated
by~\citet{Alam2019}~(see their Fig. 1), using $\afiv$ defined in this way
minimizes the residual correlation between the tidal anisotropy $\afiv$ and
the overdensity $\delta_8$, the galaxy overdensity within a sphere of
radius $5\,\hmpc$ centred on each galaxy, computed in the same way as
$\delta_5$ in Equation~\ref{eqn:afiv}. Therefore, any observed bar
dependence on $\afiv$ in our analysis should in principle be free of
contamination from the potential dependence of bars on $\delta_8$.

\section{Bar Detection Method}
\label{sec:det}

\begin{figure}
	\includegraphics[width=0.96\columnwidth]{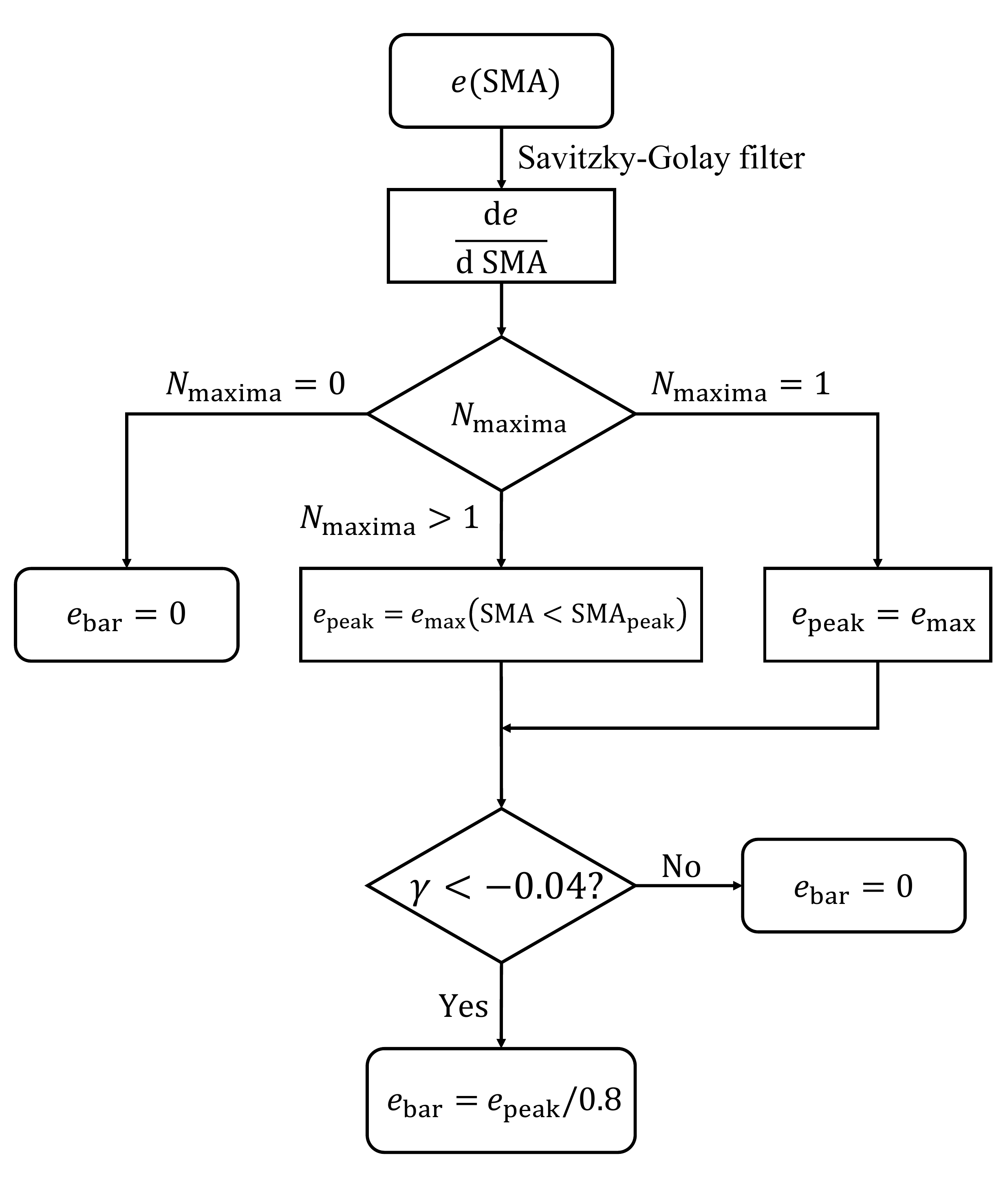}
	\caption{Flowchart of our automated bar detection algorithm based on ellipse
        fitting. We start from the ellipticity profile $e(\mathrm{SMA})$
        measured from the SDSS {\it r}-band atlas image either before or after
        the disc subtraction, and arrive at our bar
        strength estimate $e_\mathrm{bar}$. See text for details.}
        \label{fig:flowchart}
\end{figure}

To robustly characterize the barred galaxy population over a relatively
large redshift range~($z{=}0.01-0.11$) using SDSS images, we develop an
automated bar detection method based on the ellipse fitting of galaxy
isophotes. In particular, to highlight any potential bar-like structure
within the co-rotation radius, we subtract the best-fitting 2D model of the
disc component from each galaxy image before searching for bars.  Compared
with the conventional ellipse fitting, our method improves the bar
detection accuracy significantly at the high redshift by enhancing the
image contrast within the central region against the overall reduction in
image quality and spatial resolution.

In the following we describe the ellipse fitting~(\S\ref{subsec:disc_sub}),
bar identification~(\S\ref{subsec:bar}), and compare our bar detection
results with the $\pb$ from GZ2~(\S\ref{subsec:gz2}), including visual
validations using galaxy images from the DESI imaging data. Readers who are
only interested in the observational results on the tidal dependence of
bars can skip this section to the results in \S\ref{sec:tidalbias}.

\subsection{Ellipse fitting and disc subtraction}
\label{subsec:disc_sub}

We carry out isophote fitting using the Python
implementation~(\texttt{photutils.Ellipse}) of the standard iterative
ellipse fitting method of \citet[][]{Jedrzejewski1987}.  We apply the fit
twice on the atlas image stamp of each galaxy, first before and then after
subtracting the 2D disc component. During each fit, we measure the
isophotes at logarithmic intervals of semi-major axis length~(SMA),
starting from the innermost 1 arc-second
radius to two times $R_{90}$, the radius that encloses $90\%$ of the Petrosian
flux in {\it r}-band. Since each isophote is characterised by its SB level,
ellipticity, PA, and SMA, we can obtain two sets of SB, ellipticity, and PA
profiles as functions of SMA for each galaxy, one with the disc and the
other without.

\begin{figure}
	\includegraphics[width=0.96\columnwidth]{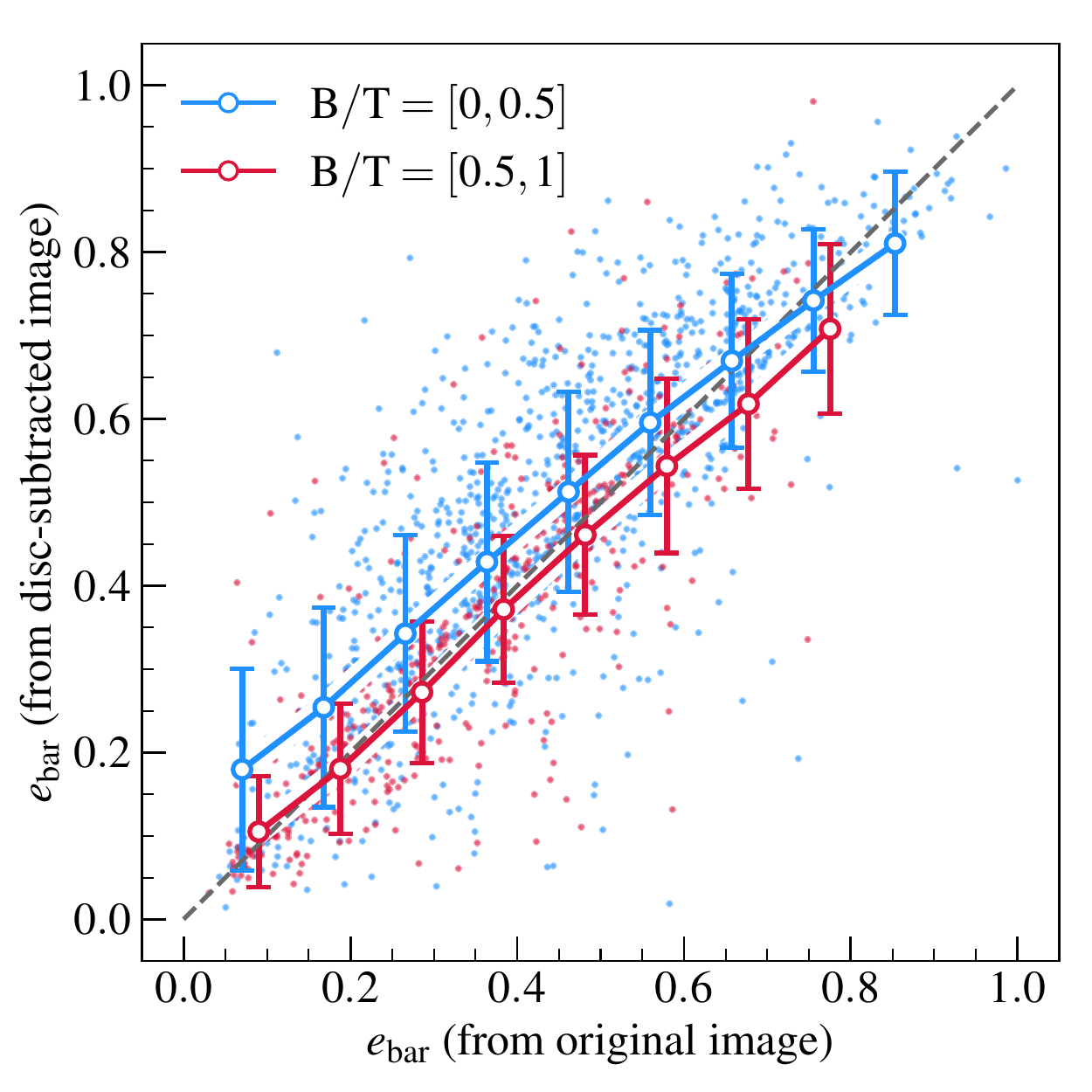}
    \caption{Comparison between the $\eb$ measurements before~(x-axis) and
    after~(y-axis) the disc subtraction. Blue and red dots indicate the
    measurements of individual galaxies with $\BT<0.5$ and $\BT\ge 0.5$,
    respectively. Blue~(Red) open circles with errorbars indicate the
    mean relation with $1{-}\sigma$ scatter for the $\BT<0.5$~($\BT\ge 0.5$) populations.} \label{fig:ebarcomp}
\end{figure}

Figure~\ref{fig:disc_subtraction_example} illustrates the efficacy of
disc-subtraction in revealing the bar embedded in the bright disc of a
face-on galaxy at $z{=}0.07$.  From left to right, we show the SDSS {\it gri}
colour composite image, the SDSS {\it r}-band image, the best-fitting 2D disc
model of \citet{Simard2011}, the SDSS {\it r}-band image after subtracting the
2D disc model, and the {\it grz} colour composite image from DESI imaging,
respectively. The bar barely shows up in the SDSS colour composite and the
{\it r}-band image, but after the 2D disc model is subtracted, the SDSS
{\it r}-band image clearly reveals a bar-like structure connected by spiral
arms at both ends, which is corroborated by the deeper image from the DESI
Legacy Surveys~\citep{Dey2019}.

\begin{figure*}
	\includegraphics[width=\textwidth]{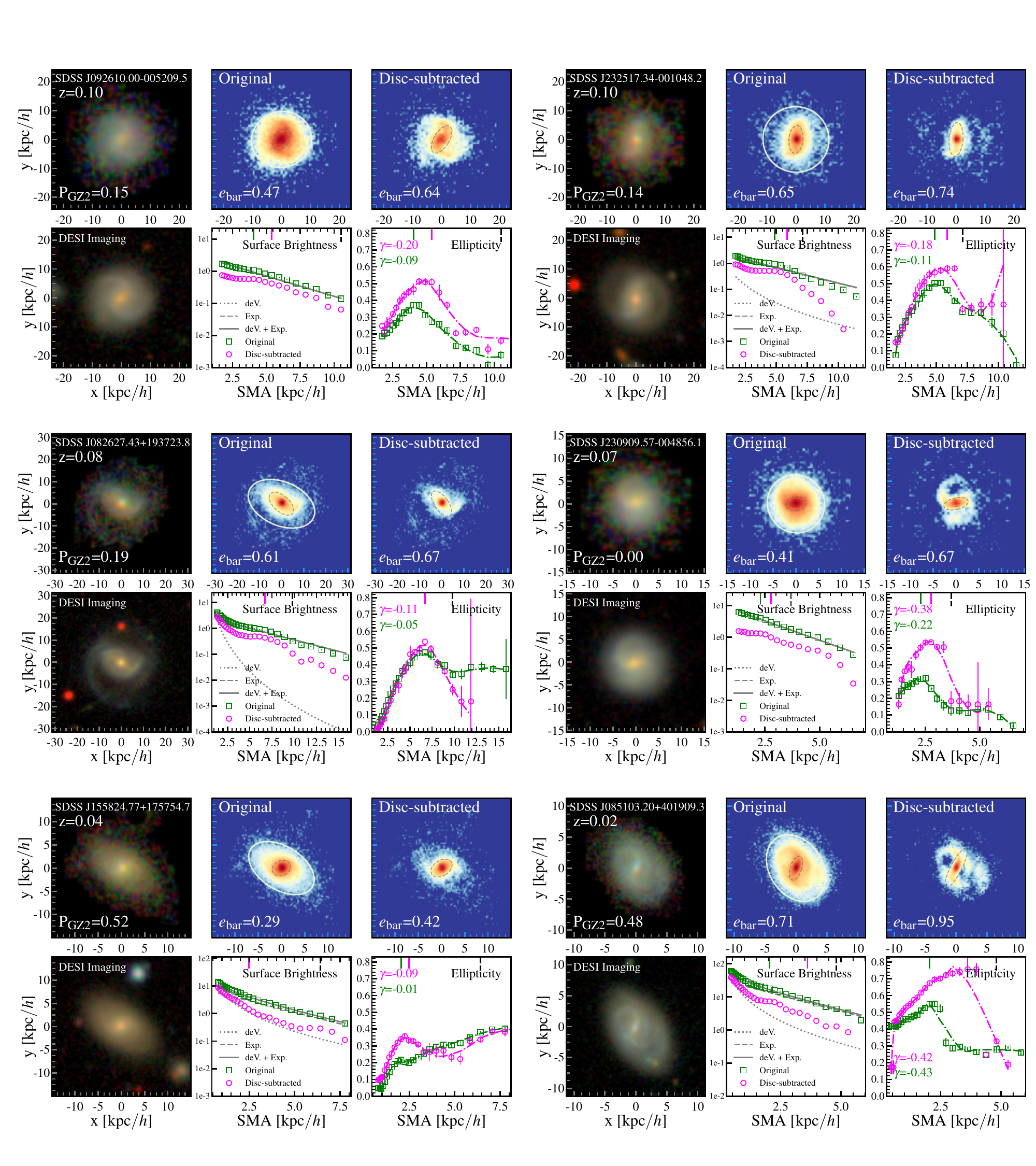}
    \caption{Demonstration of our bar detection method based on the ellipse
    fitting using six example galaxies, ordered by decreasing
    redshift~(from left and
right, top to bottom). For the three-by-two set of panels for each galaxy,
the top
    left panel shows the SDSS {\it gri} colour composite image, with the
    galaxy id and redshift listed in the top left corner and the
    GZ2-estimated bar probability $\pb$ in the bottom left corner.
The bottom left panel shows the DESI {\it grz}
    colour composite image of the same galaxy.  The top middle and right
    panels are the original and disc-subtracted SDSS {\it r}-band images,
    with the measured $\eb$ labeled in the bottom left corner,
    respectively. The gray dashed ellipses in both panels indicate the
    isophote with maximum ellipticity, while the white solid
    ellipse in the top middle panel marks the isophote that encloses $90\%$
    of the total luminosity. The bottom middle and right panels show the
    surface brightness and the ellipticity profile measured from the
    original image~(green squares) and the disc-subtracted image~(magenta
    circles), respectively. In the bottom middle panel, the solid gray
    curve is the best-fitting 1D SB model consisting of a de
    Vaucouleurs~(dotted) and an exponential~(dashed) profile. In the bottom
    right panel, the dot-dashed curves are the smoothed ellipticity profile
    derived from the Savitzky-Golay filter, and the two estimates of the
    minimum slope $\gamma$ are indicated in legend.} \label{fig:sixstamps}
\end{figure*}

To subtract the 2D disc component of a galaxy, we make use of the
best-fitting 1D and 2D SB profiles of the disc component derived by
\citet{Simard2011} using a 2D bulge-disc decomposition method.  Briefly, we
fit a combination of a $\mathrm{S\acute{e}rsic}\,n=4$ bulge and an
exponential disc 1D model to the measured SB profile to get the SB
amplitude of the disc component.  Together with the scale length,
inclination angle, and PA of the disc derived by \citet{Simard2011}, we
build a 2D SB model for the disc and subtract it from the galaxy {\it r}-band
image. In some cases, we scale down the amplitude of the disc model in
order to avoid hitting zero SB in the outskirt of the disc-subtracted
image.  After the 2D disc component is subtracted, we then re-apply the
same isophote fitting method to the disc-subtracted image to obtain our
fiducial ellipticity and PA profiles for bar detection.

\subsection{Bar identification}
\label{subsec:bar}

We search for the presence of bars based on their distinct imprint on the
ellipticity and PA profiles of galaxies. In particular, as a highly
elongated structure through the galaxy centre, a bar would induce an
increasing trend of the ellipticity profile on small scales, resulting in a
prominent peak at ${\sim}1.5{-}10\,\kpc$~\citep{Marinova2007} that then
drops off rapidly as the bar ends at close to the co-rotation
radius~\citep[e.g.,][]{Laine2002,Jogee2004,Aguerri2005,Marinova2007,Li2011}.
Meanwhile in the PA profile, the bar should maintain a relatively constant
PA for the isophotes inside the bar region, before transitioning to the PA
of the outer disc. For a real stellar bar, the peak ellipticity is closely
related to the eccentricity of the periodic orbits in the $x1$ family that
underpins the bar, hence providing us a robust measure of the bar
strength~\citep{Athanassoula1992,Martin1995,Marinova2007,Li2011}.

Following the philosophy outlined above, our automated bar detection
algorithm is illustrated by the flowchart shown in
Figure~\ref{fig:flowchart}. Given the observed ellipticity profile of a
face-on disc galaxy, we calculate its first-order derivative profile
$\mathrm{d}e/\mathrm{d}\mathrm{SMA}$ using the \citeauthor{Savitzky1964}
filter that accounts for the uncertainties of the ellipticity
measurements\footnote{\url{https://github.com/surhudm/savitzky_golay_with_errors}.
We set the \texttt{window\_length}{=}7 and \texttt{degree}{=}3 during the
smoothing.  }.  We then identify all the $N_{\mathrm{maxima}}$ local maxima
between $1.5\,\kpc$ and $\mathrm{SMA}_{90}$ based on the derivative
profile.  The minimum search radius is chosen to exclude the possible
impact from a bulge, while for galaxies with noisy ellipticity profiles we
modify the maximum search radius to the SMA where the error in ellipticity
exceeds $0.05$ for at least three consecutive isophotes.

If no peaks are found~($N_{\mathrm{maxima}}{=}0$), we identify the galaxy as
barless and assign it a zero $\epk$, and if $N_{\mathrm{maxima}}{=}1$, we
assign the ellipticity of that single peak to the galaxy as its $\epk$.
However, in some cases, there are more than one peak and the highest peak
usually corresponds to the ellipticity of the outer region rather than
the bar. To correctly identify the bar in the inner region, we assign the
ellipticity of the secondary peak found within the radius of the highest
peak~(but still above $1.5\,\kpc$) as the $\epk$. We mark the location of the
peak ellipticity as $\rpk$. After this step, we identify $92814$ out of the
$100445$ face-on galaxies as having positive $\epk$.

Next, we quantify the steepness of the decline following the peak
ellipticity using the minimum slope parameter $\gamma$, defined as
\begin{equation}
    \gamma = \mathrm{min}\left( \frac{\mathrm{d}\, e}{\mathrm{d}\,
    \mathrm{SMA}} \right)_{\sma > \rpk}.
	\label{eqn:gamma}
\end{equation}
Since $\gamma$ is negative, the absolute value of $\gamma$ is larger when
the elongated structure ends more abruptly, hence a higher probability of
being a real bar. Some inclined discs or bulge-dominated galaxies may
exhibit a peak in the ellipticity profile despite the lack of a bar. In
those systems, the peak would slowly give way to the ellipticity of the
outer light distribution, rather than experiencing a sharp cutoff. To
remove those false positives, we set the peak ellipticity of any galaxy
with $\gamma{>}{-}0.04$ to be zero.  We have verified that our results are
insensitive to the choice of the minimum value of $\gamma$~(at least
between ${-}0.08$ and ${-}0.02$).  This procedure reclassifies $29312$ out
of $92814$ galaxies with positive $\epk$ into the barless category, with
$63502$ galaxies remained possibly barred.

Following \citetalias{Yoon2019}, we require the variation of PA to be less
than $20^{\circ}$, starting from where the ellipticity first reaches $0.25$
to $\rpk$. The stable PA requirement helps excluding the high ellipticities
caused by spurious features such as some tightly wound spiral arms. We do
not require a change in the PA profile between the bar and the disc as the
disc is subtracted in our fiducial analysis. The PA selection further
removes $2009$, leaving $61493$ barred galaxies with $\epk{>}0$.

Finally, we can measure two types of bar strength for each galaxy using the
$\epk$ measured from the original and disc-subtracted images. For the sake
of convenience, we define a new bar ellipticity parameter $e_\mathrm{bar}$
by normalising the values of $\epk$
\begin{equation}
    e_\mathrm{bar}=\frac{e_\mathrm{peak}}{\mathrm{max}(e_\mathrm{peak})},
\end{equation}
where the denominator is the maximum peak ellipticity, which is $0.8$ in
either set of measurements. We will use $\eb$ to quantify bar strength
throughout the rest of the paper.

\begin{figure}
	\includegraphics[width=0.96\columnwidth]{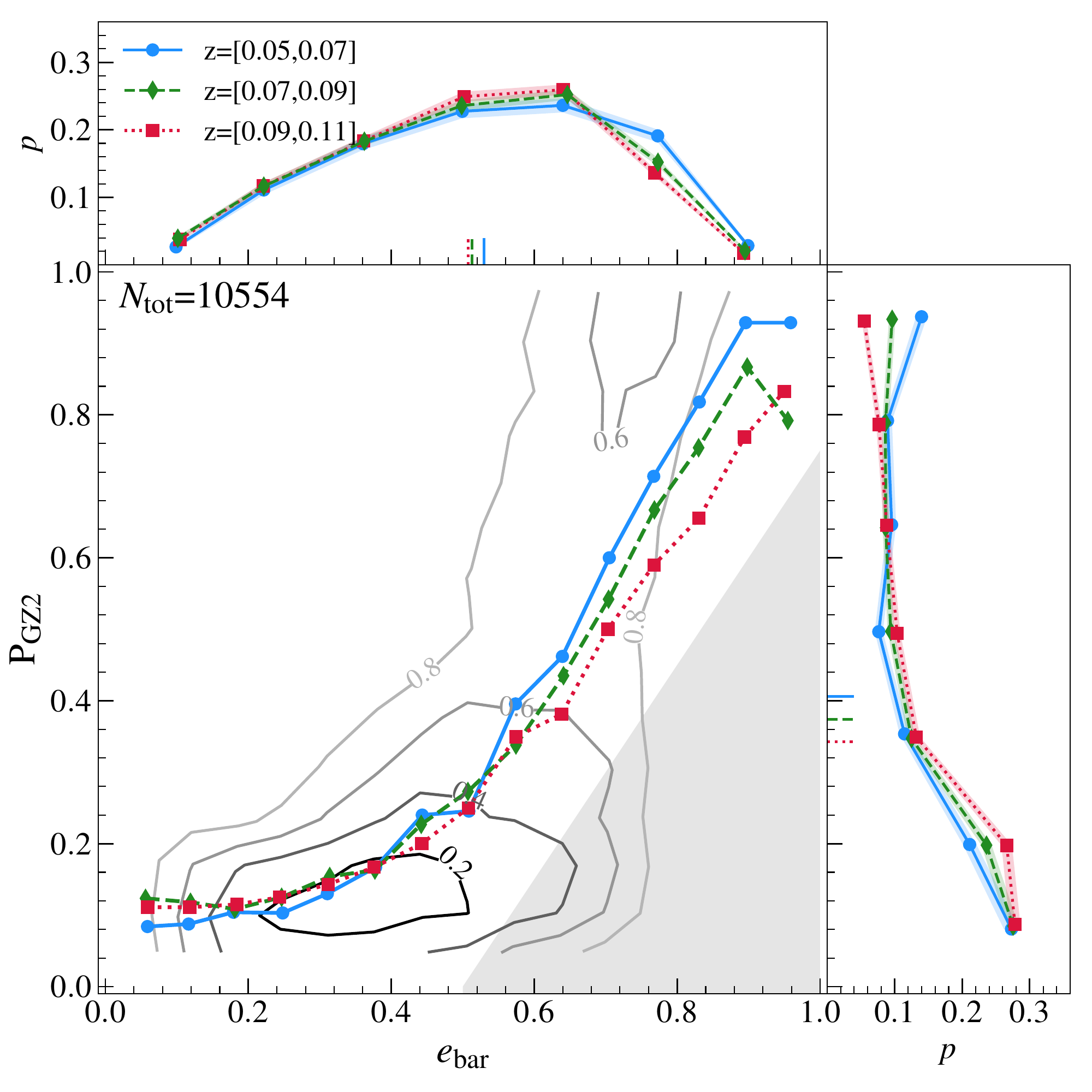}
    \caption{Distribution of barred galaxy candidates~(i.e., with both
    $\eb{>}0$ and $\pb{>}0$) from
    the luminosity-limited sample on the $\pb$ versus $\eb$ plane. In the
    main panel, the contour
    levels represent $20\%$, $40\%$, $60\%$, and $80\%$ of the galaxies,
    with the total number of galaxies listed in the top left corner. The
    curves of different colours and symbols show the median relationships
    in three redshift slices~(blue solid for $0.05{<}z{<}0.07$, green
    dashed for $0.07{<}z{<}0.09$ and red dotted for $0.09{<}z{<}0.11$).
    The shaded triangular region in the bottom right corner highlights the
    galaxies with strong discrepancies between the two measurements.  The
    two side panels show the 1D probability distributions and the
    corresponding errors~(shaded) of $\eb$~(top) and $\pb$~(right) for the
    three redshift slices, colour-coded the same way as in the main panel.
    The short vertical rungs of matched colours and styles indicate the
    mean values of the 1D distributions.}
	\label{fig:ebarvGZ}
\end{figure}

Figure~\ref{fig:ebarcomp} compares the two $\eb$ measurements before and
after the disc subtraction, with each blue~(red) dot representing a galaxy
with $\BT{<}0.5$~($\BT{\ge}0.5$). The blue~(red) open circles with
errorbars indicate the mean relation between the two measurements along
with its $1{-}\sigma$ scatter for the $\BT{<}0.5$~($\BT{\ge}0.5$) galaxies.
As mentioned in the introduction, the measured shape of the bar can be
blunted by the light from the disc~\citep{Gadotti2008}, leading to an
underestimation of the bar ellipticity or even an undetected bar.  Compared
with the one-to-one relation~(dashed line), the mean relation of the
disk-dominated galaxies~(blue circles) indicates that after disc
subtraction, $\eb$ generally increases by an amount between $0.1$ and
$0.05$ for galaxies with weak bar-like features~($\eb{<}0.4$), and stays
roughly unchanged for galaxies with strong bars~($\eb{>}0.7$).  This
behaviour before and after the disc subtraction is consistent with our
expectation that the disc subtraction helps mitigating the impact from
bright discs in weakly-barred systems. Meanwhile, the mean relation of the
bulge-dominated galaxies~(red circles) indicates that the two measurements
are roughly consistent for most of the galaxies, but exhibit a slight
decrease in $\eb$ after the disc subtraction for those with $\eb{>}0.7$.
Since many of the high-$\eb$, high-$\BT$ systems are false positives, the
decrease is probably caused by the fact that the bulge component becomes
less bar-like after the disk removal, hence an improvement of the method.
In addition, we have performed a suite of mock tests of the impact of
inaccurate 2D disc models on the bar strength, and find that the number of
false positives caused by the small uncertainties associated with the
\citet{Simard2011} disc models is negligible in our paper.  As a result, we
adopt the $\epk$ measured from the disc-subtracted SB profiles as our
fiducial peak ellipticity estimates.

\begin{figure}
	\includegraphics[width=0.96\columnwidth]{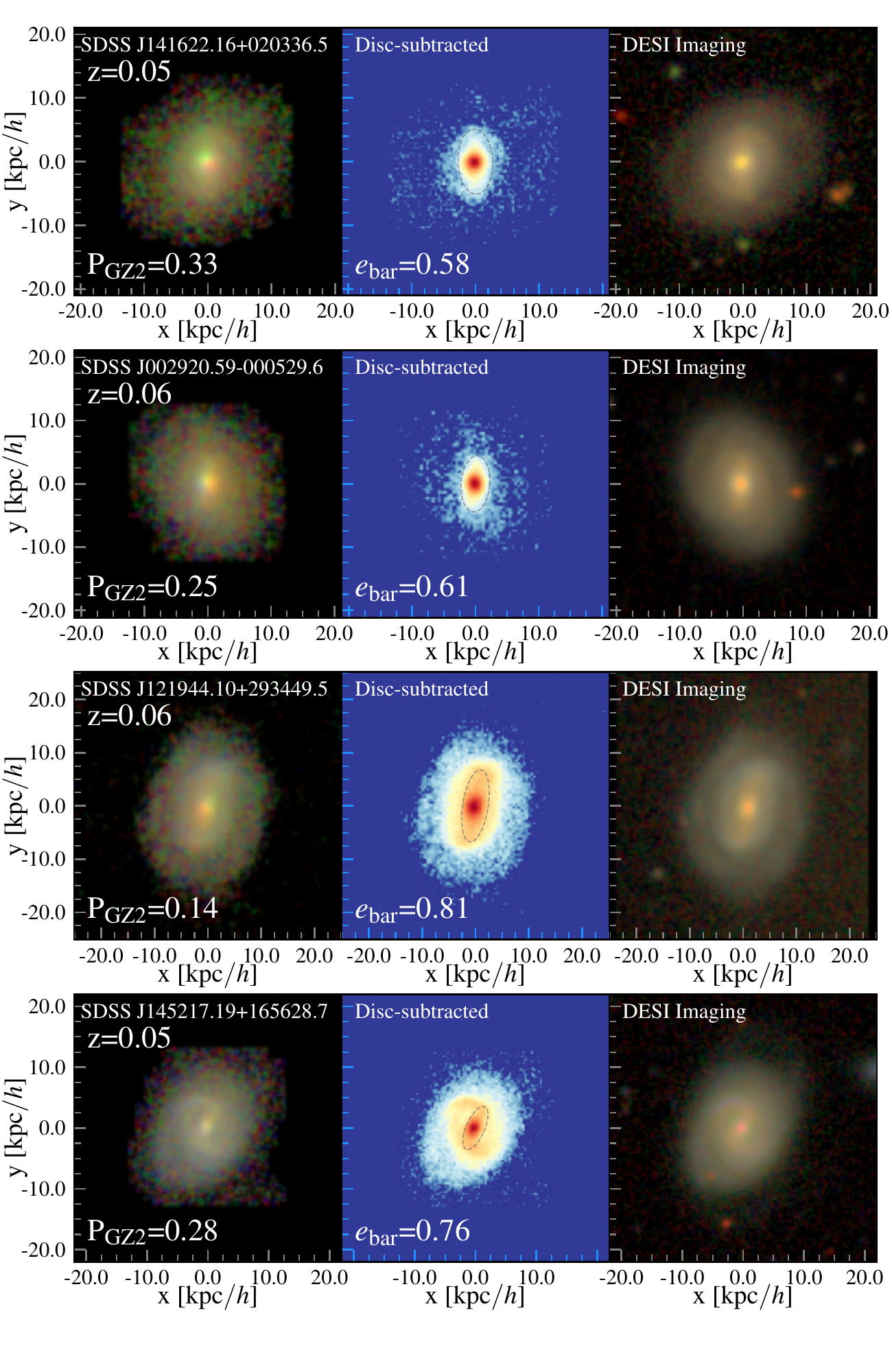}
    \caption{Examples of four galaxies with high $\eb$ but low $\pb$. In
    each row, the left panel shows the SDSS {\it gri} colour composite
    image of the galaxy, with $\pb$ listed in the lower left corner. The
    middle panel shows the disc-subtracted SDSS {\it r}-band image along
    with the ellipse with maximum ellipticity~(gray dashed) and the
    measured $\eb$ value in the lower left. For comparison, the right panel
    shows the DESI {\it grz} colour composite image of the same galaxy.}
    \label{fig:eoverp_stamp}
\end{figure}

Figure~\ref{fig:sixstamps} demonstrates the efficacy of our bar detection
method based on the ellipse fitting of disc-subtracted images of six
candidate barred galaxies, ordered by decreasing redshift~(from left and
right, top to bottom).  For the three-by-two panels of each galaxy, the top
left panel is the SDSS {\it gri} colour composite image, with the galaxy id and
redshift listed in the top left corner and the GZ2-estimated bar probability
$\pb$ in the bottom left corner.  The top middle and top right panels
display the original and disc-subtracted SDSS {\it r}-band images of the
galaxy, respectively. In each of the two panels, the dashed gray ellipse
indicates the isophote with maximum ellipticity, which corresponds to the
location of the candidate bar~($\eb$ value listed in the bottom left
corner).  The solid white ellipse in the top middle panel indicates the
isophote that encloses $90\%$ of the total luminosity, hence with
ellipticity $e_\mathrm{90}$.  The bottom left panel displays the DESI {\it grz}
colour composite image of the same galaxy, from which we can better
evaluate the ellipticity and PA of the bar-like structure visually.  The
bottom middle and right panels show the SB and ellipticity profiles,
respectively.  In the bottom middle panel, the green squares with errorbars
are the 1D SB profile measured from the original SDSS {\it r}-band image shown
in the top middle panel. The gray solid curve indicates the best-fitting
1D SB model from~\citet{Simard2011} to the green squares, consisting of a
de Vaucouleurs component~(dotted) and an exponential disc
profile~(dashed).  The magenta squares with errorbars indicate the 1D SB
profile measured from the disc-subtracted {\it r}-band image shown in the top
right panel.  The bottom right panel compares the ellipticity profile
measured from the disc-subtracted image~(magenta circles with errorbars) to
that from the original image~(green squares with errorbars), each fitted with
a smooth model~(dot-dashed curves) derived from the Savitzky-Golay filter.
The two estimates of the minimum slopes are also listed in the top left
corner.

In the first four cases shown in Figure~\ref{fig:sixstamps}, the visual
evidence of a bar presence in the SDSS images is rather weak, consistent
with the low $\pb$ values given by GZ2~(below $0.2$).  However, the DESI
images generally reveal a much stronger bar-like structure in the centre
than the SDSS images, suggesting that the lack of visual detections in GZ2
is due to the fact that those four galaxies are relatively distant with
redshifts $z{>}0.07$. Using the SDSS original {\it r}-band images, our ellipse
fitting method successfully identifies an elongated structure in each
galaxy, with $\eb$ values ranging from $0.41$ to $0.65$~(thick green
ellipses), but the PAs of the ellipses are generally offset from the actual
bar shown in the DESI images by $5{-}15$ degrees.  Such a PA offset is not
limited to the four high-redshift galaxies, but exists even in the bottom
row where the two galaxies are relatively nearby with redshifts below
$0.04$, signaling a bias in the $\eb$ derived using the original SDSS
images regardless of redshift.

\begin{figure*}
	\includegraphics[width=0.96\textwidth]{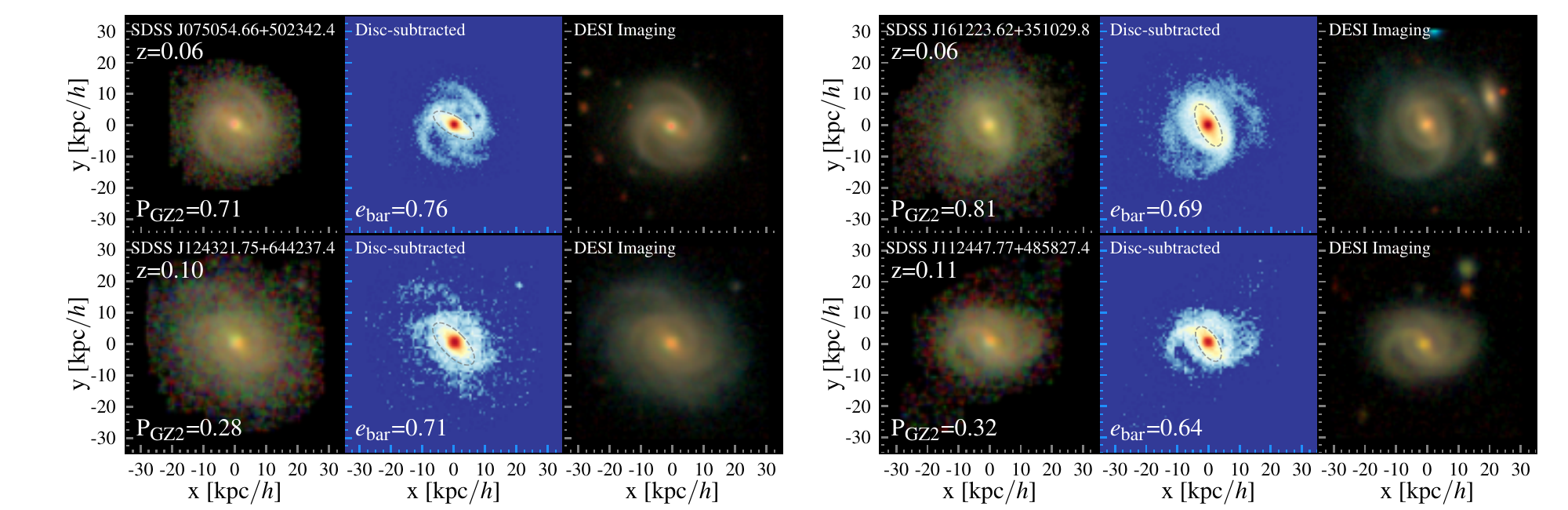}
    \caption{Two pairs of "twins" galaxies that have similar physical and
    morphological properties but observed at different redshifts. For each
    pair in the same column, the top row shows the standard three-panel
    view~(same format as in Figure \ref{fig:eoverp_stamp}) of the galaxy at
    the lower redshift, while the bottom row shows its "twin" at the higher
    redshift. While the $\eb$ and $\pb$ measurements agree well at the low
    redshifts, they tend to differ significantly at the higher redshifts.}
	\label{fig:hlz_stamps}
\end{figure*}

The PA offset problem is significantly alleviated in our fiducial bar
detection method using disc-subtracted images, suggesting the bar
morphologies are better approximated by the highest-$\epk$ ellipses in our
fiducial measurements.  As a result, the ellipticity profiles are generally
more strongly peaked after disc subtraction compared to the original ones,
yielding increases in $\eb$ ranging between $\Delta\eb{=}0.05$ and $0.23$.
In addition, the minimum slope $\gamma$ generally drops after disc
subtraction~(except for the nearest object), illustrated by the strong
discontinuities in the ellipticity profiles~(magenta circles) following the
occurrences of peak ellipticities.

\subsection{Comparison with Visual Bar Identifications}
\label{subsec:gz2}

We now quantitatively compare the bar strengths $\eb$ calculated by our
fiducial bar detection method with the bar probabilities $\pb$ derived by
the state-of-the-art visual identifications from the GZ2 catalogue.  To
avoid any discrepancies caused by small number statistics in the visual
inspection, we limit our comparison to galaxies that have received more
than five responses to ``whether the galaxy is barred'' by the citizen scientists in GZ2.  In total, we
have $20979$ face-on galaxies between $0.05{<}z{<}0.11$ with robust
measurements of both $\eb$ and $\pb$.  Among those, $10554$ of them are
measured with both $\eb{>}0$ and $\pb{>}0$, $4117$ with $\eb{=}0$ but $\pb{>}0$,
and $2733$ with $\eb{>}0$ but $\pb{=}0$. By carefully examining the DESI images
of those galaxies in the latter two categories, we find that
the success rate of either bar detection method is close to $50\%$,
suggesting the two methods are comparable in identifying the truly barless
galaxies. Below we will focus on the $10554$ galaxies with both positive
values of $\eb$ and $\pb$, as many of them are likely truly barred galaxies
that are suitable for comparing the two types of bar strength estimates.

Figure~\ref{fig:ebarvGZ} compares the overall 2D distribution~(contours)
and median relationships in three different redshift bins~(blue solid:
$0.05{<}z{<}0.07$; green dashed: $0.07{<}z{<}0.09$; red dotted:
$0.09{<}z{<}0.11$) on the $\eb$ vs. $\pb$ plane in the main panel, with
their respective 1D distributions in three redshift bins shown in the two
side panels.  Overall, there exists a good correlation between $\pb$ and
$\eb$ with a Pearson cross-correlation coefficient of $0.57$,
indicating a reasonable agreement between the GZ2 visual inspection and our
ellipse fitting methods. However, the median $\pb$ as a function of $\eb$
at fixed redshift is not a diagonal one-to-one relation, but appears flat
at $\eb {<} 0.4$ before rising steeply at $\eb {\ge} 0.4$.  The flattening is
associated with the large number of low-$\pb$ galaxies clustered below
$\pb{=}0.3$, as shown by the 1D PDFs of $\pb$ in right panel.  Despite being
confined within $\pb{=}0.3$, those low-$\pb$ galaxies have a wide spread in
$\eb$, with a good portion of them identified as strongly barred~($\eb {\sim}
0.8$) by our method.  The slope of the median relationship at the
high-$\eb$ end evolves significantly with redshift, largely due to the
decrease of high-$\pb$ galaxies with increasing redshift~(right panel).
Meanwhile, the 1D PDF of $\eb$ is approximately redshift-independent,
exhibiting a single broad peak at $\eb{\sim} 0.6$.  Therefore, assuming that
the redshift evolution of the bar fraction is negligible across the narrow
redshift range, Figure~\ref{fig:ebarvGZ} suggests that the $\eb$ estimates
are relatively insensitive to the decrease of both the signal-to-noise and
physical resolution of the galaxy images with redshift.

To investigate the origin of discrepancies between the two methods, we
select all the $1213$ galaxies from the high-$\eb$, low-$\pb$ corner in
Figure~\ref{fig:ebarvGZ}~(shaded triangular region) and visually compare
their corresponding SDSS vs. DESI images. Unsurprisingly, the SDSS images
of these galaxies barely exhibit any signatures of bars, but their DESI
images tell a very different story --- at least $52\%$ of these galaxies
are strongly barred based on the DESI images and another $34\%$ of them are
likely barred with oval distortions in the centre.
Figure~\ref{fig:eoverp_stamp} shows four examples of such high-$\eb$,
low-$\pb$ galaxies from the lowest redshift bin~($0.05{<}z{<}0.07$), where the
physical resolutions of SDSS and DESI images should both be adequate for
resolving the strong bars. The four galaxies show little signature of
having a bar in the SDSS images~(left column), but appear strongly barred
based on the values of $\eb$~(middle column) and the DESI images~(right
column).  The formats of individual panels in each row are the same as the
respective panels in Figure~\ref{fig:sixstamps}. In general, the discs of
these galaxies are relatively bright, rendering the bar less prominent in
the SDSS images; Our ellipse fitting over the disc-subtracted images is
able to identify the correct ellipses~(gray dashed ellipses) that match the
shape and extent of the bars seen in the DESI images very well.

\begin{figure}
	\includegraphics[width=0.96\columnwidth]{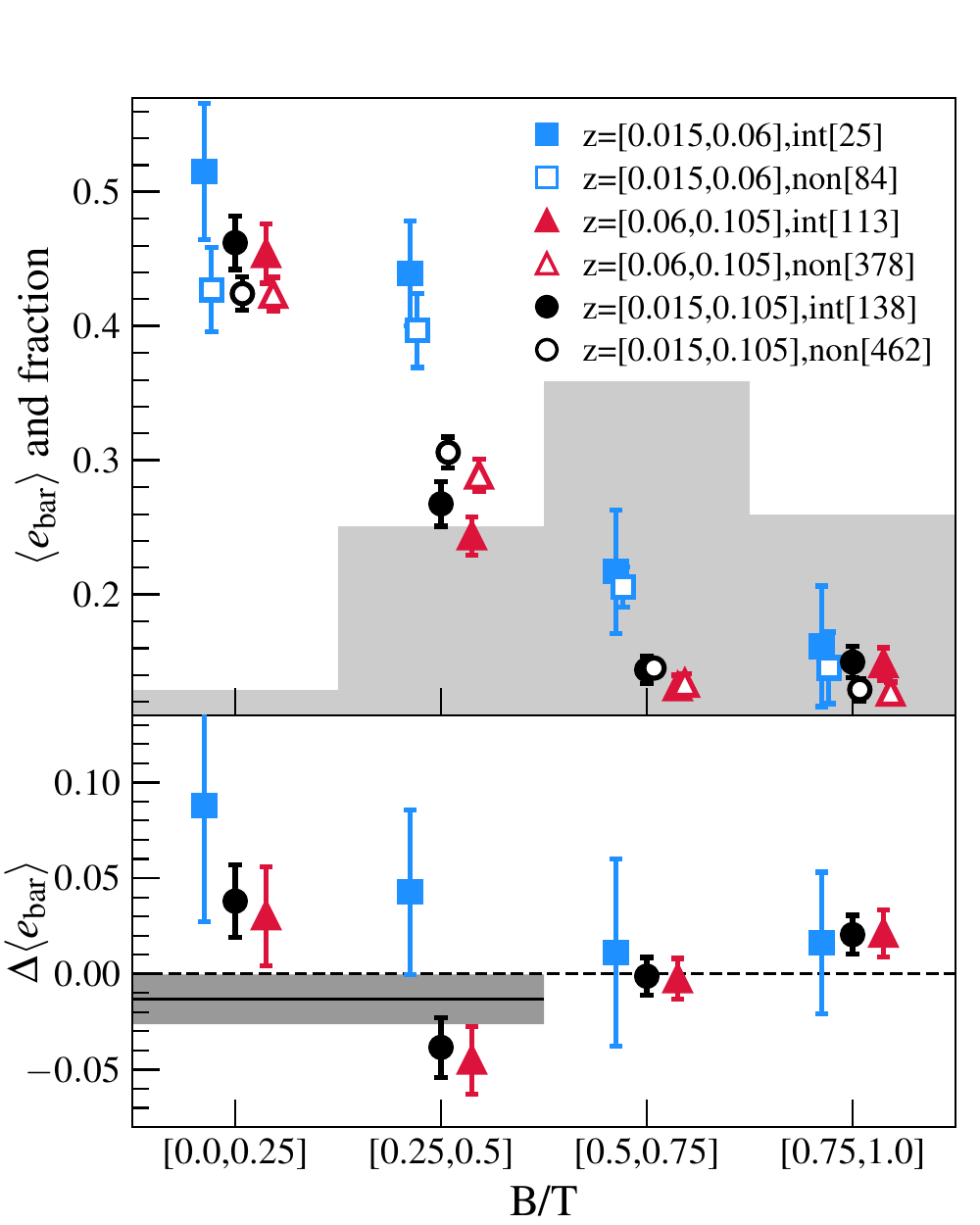}
    \caption{The mean bar strength $\langle\eb\rangle$ of member galaxies
    with different $\BT$ in interacting~(solid symbols) vs.
    non-interacting clusters~(open symbols), while the differences in their
    $\langle\eb\rangle$ between interacting and non-interacting clusters
    are shown in the bottom sub-panel.  Blue squares, red triangles, and
    black circles are for the low-redshift~($0.015{<}z{<}0.06$),
    high-redshift~($0.06{<}z{<}0.105$), and the full-redshift range cluster
    samples, respectively, with the numbers of different type of clusters
    in each redshift bin listed in the legend.  The light-shaded histogram
    in the top panel indicates the relative abundance of member galaxies in
    the four different $\BT$ bins.  The black horizontal band in the bottom
    panel~($\Delta\meb{=}{-}0.013{\pm}0.013$) indicates the overall discrepancy between interacting and
    non-interacting clusters for galaxies with $0{<}\BT{<}0.5$.  All the
    errorbars are the $1-\sigma$ uncertainties estimated from Jackknife
    resampling.}
	\label{fig:yoon}
\end{figure}

Using the disc-subtracted galaxy images, our automated bar detection method
is relatively insensitive to the image quality and physical resolution, as
demonstrated by the lack of redshift evolution in the 1D PDFs of $\eb$ in
the top panel of Figure~\ref{fig:ebarvGZ}.  This is very encouraging,
showing that it is viable to extend the SDSS analysis of tidal dependence
of bars from the local Universe at $z{<}0.06$~(as was done in
\citetalias{Yoon2019}) to a much larger volume up to $z{=}0.11$ with five
times more galaxy clusters.  Figure~\ref{fig:hlz_stamps} further
illustrates the robustness of our fiducial method against redshift with two
pairs of barred galaxies. For each pair in the same column, the top row
shows the standard three-panel view~(same format as in
Figure~\ref{fig:eoverp_stamp}) of one galaxy at a lower redshift, which can
be compared to the bottom row in which we show the same view of its
``twin'' galaxy with very similar physical properties~({\it r}-band luminosity
and effective radius) and appearance~(morphology and inclination angle
based on the DESI images) but at a higher redshift. For the two galaxies at
$z{=}0.06$~(top row), both the GZ2 and our ellipse fitting method regard them
as strongly barred, with $\pb{=}(0.71, 0.81)$ and $\eb{=}(0.76, 0.81)$,
respectively. As expected, the SDSS image quality decreases considerably at
$z{\ge} 0.10$, resulting in much lower values of $\pb$ for the two ``twin''
galaxies in the bottom row~($\pb{=}0.28, 0.32$). However, our bar detection
method yields $\eb$ values~($\eb{=}0.71, 0.64$) that are consistent with
their low-$z$ counterparts~($\eb{=}0.76, 0.69$), confirming our expectation
based on the DESI images.

\begin{figure*}
	\includegraphics[width=0.96\textwidth]{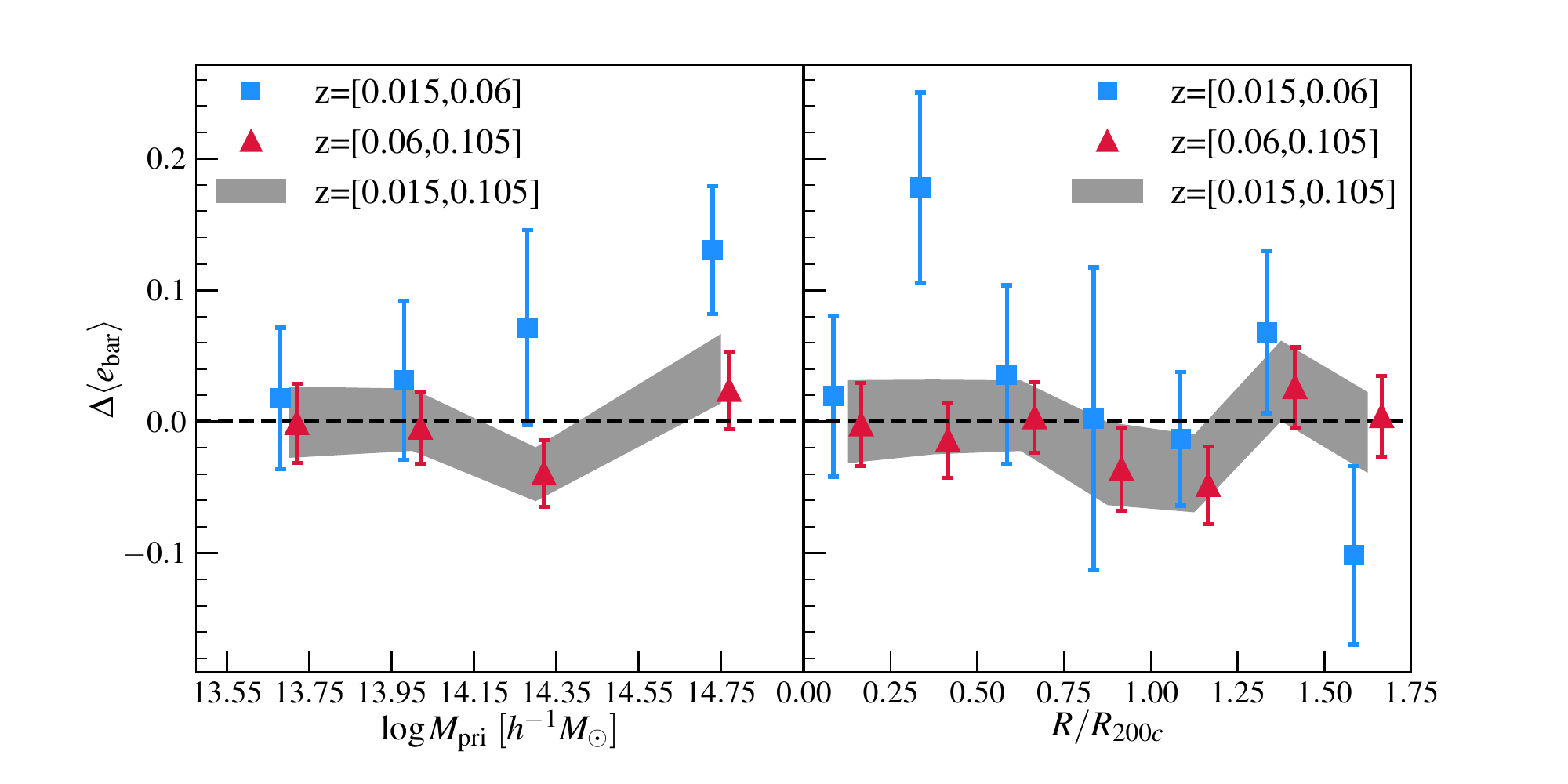}
    \caption{The dependence of $\Delta\meb$ on
    the mass of the primary cluster $M_\mathrm{pri}$~(left panel) and the
    projected cluster-centric distance scaled by the halo radius,
    $R/R_{200c}$~(right panel). Blue squares and red triangles are for the
    low~($0.015{<}z{<}0.06$) and high~($0.06{<}z{<}0.105$) redshift subsamples,
    respectively. The dark shaded bands are the final results from the
    overall cluster sample. All errorbars are $1{-}\sigma$ uncertainties
    estimated from Jackknife resampling.}
	\label{fig:evMpSh}
\end{figure*}

To summarise, our fiducial bar detection method based on the ellipse
fitting, when applied to the SDSS {\it r}-band images of galaxies with their
disc components subtracted, is capable of providing robust measurements of
the bar strength parameter $\eb$. Overall, our method is largely consistent
with the visual identification results like the GZ2 at the low redshifts.
At higher redshifts, our fiducial bar detection remains robust against the
reduction of image quality with redshift at least up to $z{=}0.11$, the
maximum redshift of our analysis in the next section.

\section{Tidal Dependence of Bars}
\label{sec:tidalbias}

Equipped with a robust bar detection method, we are now ready to examine
the dependence of bar strength, characterised by $\eb$, on the different
tidal environments measured in \S\ref{sec:tidal}.  In particular, we
examine the tidal dependence of bar strength $\eb$ on cluster scales up to
three times the virial radius in \S\ref{subsec:cluster_scale}, and then
shift our focus to the tidal anisotropy field defined over $5\,\hmpc$
scales~(measured by $\afiv$) in \S\ref{subsec:large_scale}. Theoretically,
we expect the small and/or large-scale tidal density fields to both
correlate with halo spin, so that any potential tidal dependence of bars
may be an indirect evidence of bar dependence on halo spin. In practice, however,
our investigation is an agnostic probe of tidal dependence of bars
regardless of the physical origin.

\subsection{Bar Dependence on the Cluster-scale Tidal Environment}
\label{subsec:cluster_scale}

\subsubsection{Is There a Boost in the Bar Strength Surrounding Interacting
Clusters?}
\label{subsubsec:boost}

For characterising the cluster-scale tidal environment, we follow the
approach of \citetalias{Yoon2019} in \S\ref{subsec:clustersplit} and split
the \citetalias{Yang2007} cluster sample into two subsamples of interacting
vs. non-interacting clusters. The strong tidal forces generated during
cluster-cluster interactions could induce ordered shear flows and spin up
haloes in the vicinity of the interacting pairs, potentially leaving an
imprint on the galactic bars. Recently, \citetalias{Yoon2019} claimed the
detection of such a signal by comparing the barred fraction between
galaxies in the interacting and non-interacting clusters, but within a
relatively short redshift range of $0.015{<}z{<}0.06$. Among the $105$ clusters
in their sample, they found a statistically significant enhancement of
galaxy bar fraction surrounding the $16$ interacting clusters compared to
the $89$ isolated ones.

Compared to the \citetalias{Yoon2019} analysis, our study is different in
four major aspects.  Firstly, although the two bar detection methods are
both based on ellipse fitting, our method measures the bar strength $\eb$
using the disc-subtracted images of galaxies; Secondly, instead of using
the bar fraction based on a binary classification, we compute the average
bar strength $\meb$ of galaxies in different tidal environments, without
resorting to an arbitrary $\eb$ for dividing barred vs. non-barred
galaxies; Thirdly, our measurement uncertainties are computed using the
Jackknife resampling technique. Briefly, for each cluster sample with $N$
pairs of interacting clusters, we construct $N$ Jackknife subsamples by
removing one pair of interacting clusters and $1/N$ of the isolating
systems at a time, and estimate the uncertainty on $\meb$ as the standard
deviation of the $N$ Jackknife measurements multiplied by $\sqrt{N-1}$.
Finally and most importantly, we use the \citetalias{Yang2007} halo-based
group catalogue and extend the maximum redshift of investigation to
$z{=}0.11$, resulting in a factor of five increase in the survey volume. In
particular, while the size of our cluster sample~($109$ above $\log
M_{200c}{=}13.85$) is similar to that of \citetalias{Yoon2019}~($105$) at
$0.015{<}z{<}0.06$, our analysis includes $491$ more clusters with $\log
M_{200c}{>}13.85$ at $0.06{<}z{<}0.105$, thereby increasing the sample size by a
factor of five compared to \citetalias{Yoon2019}.

Figure~\ref{fig:yoon} compares the mean $\eb$ of galaxies between the
interacting~(filled symbols) and non-interacting clusters~(open symbols) in
four different bins of bulge-to-total ratio $\BT$. In the top
panel, blue squares and red triangles show the measurements for the clusters in
the low~($0.015{<}z{<}0.06$) and high~($0.06{<}z{<}0.105$) redshift bins,
respectively, and the combined results from the full cluster sample are
shown by the black symbols. The underlying light-shaded histograms indicate
the relative abundance of galaxies in the four different $\BT$ bins.
Filled symbols of the matching colours and styles in the bottom panel
indicate the average difference between the bar strength of galaxies around
the interacting and non-interacting clusters, defined as
\begin{equation}
\Delta \meb \equiv \langle e_{\mathrm{bar}}^{\mathrm{int}} \rangle - \langle
e_{\mathrm{bar}}^{\mathrm{non}}\rangle.
\label{eqn:delta_eb}
\end{equation}

All the errorbars are the $1-\sigma$ uncertainties estimated from Jackknife
resampling.  As expected, the disc-dominated galaxies~($\BT{<}0.5$) are more
likely to have strong bars than the bulge-dominated systems~($\BT{\ge} 0.5$)
regardless of redshift or tidal environments. The average bar strength of
galaxies with $0.25{<}\BT{<}0.75$ is significantly higher in the low redshift
bin than in the high redshift one, echoing the finding in
Figure~\ref{fig:ebarvGZ} where the fraction of high-$\eb$ galaxies is
enhanced at the low redshift.

Comparing the average bar strengths between galaxies in the interacting vs.
non-interacting clusters in the low redshift bin, we confirm the results
from \citetalias{Yoon2019}, finding that the $\meb$ of disc-dominated
galaxies with $0{<}\BT{<}0.25$~($0.25{<}\BT{<}0.5$) surrounding the interacting
clusters is $0.088{\pm}0.061$~($0.043{\pm}0.043$) higher than that around the
isolated clusters~(blue filled squares in the bottom panel of
Figure~\ref{fig:yoon}). The discrepancy is consistent with zero for
bulge-dominated galaxies with $\BT{\ge} 0.5$.  We note that the volume covered
by our low redshift bin overlaps completely with that analyzed by
\citetalias{Yoon2019}, yielding $91$ clusters in common between the two
analyses. Therefore, it is reassuring and unsurprising that the two sets of
measurements at $0.015{<}z{<}0.06$ are consistent with each other, despite the
differences in methodologies.

\begin{figure*}
	\includegraphics[width=\textwidth]{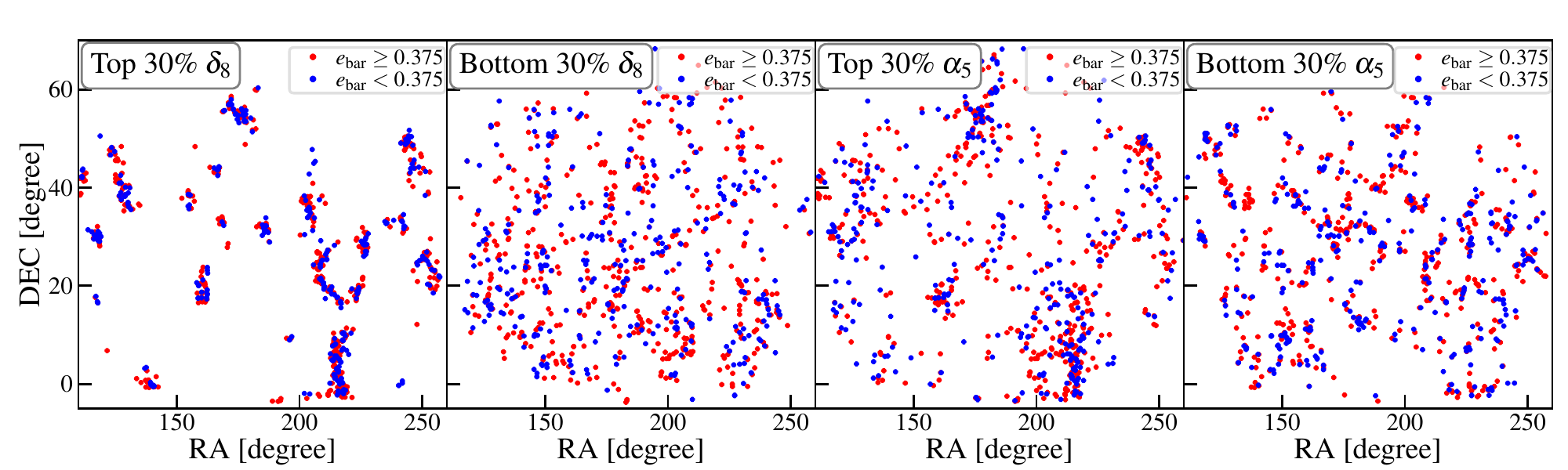}
    \caption{The spatial distribution of more-likely barred galaxies~($\eb
    {\ge} 0.375$, red) and less-likely barred galaxies~($\eb {<} 0.375$, blue)
    in the top/bottom $30\%$ of $\degt$~(left two panels) and
    $\afiv$~(right two panels) within a redshift slice of $\Delta z{=}0.007$
    centred at $z{=}0.057$.}
	\label{fig:skyscatter}
\end{figure*}

However, the discrepancy observed for the disc-dominated galaxies becomes
significantly weaker in the high-redshift bin where our statistical
uncertainties are $2{-}3$ times smaller~(red filled triangles in the bottom
panel of Figure~\ref{fig:yoon}). Intriguingly, $\Delta\meb$ remains
somewhat positive~($\Delta\meb{=}0.030{\pm}0.026$) for the pure discs with
$0{<}\BT{<}0.25$, but becomes negative~($\Delta\meb{=}{-}0.045{\pm}0.018$) for the
disc galaxies with $0.25{<}\BT{<}0.5$.  The discrepancy for the bulge-dominated
galaxies~($\BT>0.5$) remains largely consistent with zero in the
high-redshift bin. Combining the two redshift bins, the overall discrepancy
between the interacting vs.  non-interacting clusters for galaxies with
$0{<}\BT{<}0.5$ is $\Delta\meb{=}{-}0.013{\pm}0.013$~(gray horizontal band in the
bottom panel of Figure~\ref{fig:yoon}). Therefore, our analysis over the
full redshift range $0.06{<}z{<}0.105$ does not provide any evidence for the
enhancement of average bar strength surrounding the interacting clusters
compared to the isolated systems.

Focusing on the disc-dominated galaxies with $0<\BT<0.5$, we explore
whether their $\meb$ discrepancy~(or lack thereof) between interacting vs.
non-interacting clusters depends on the cluster mass or their projected
clustercentric distance in Figure~\ref{fig:evMpSh}.  The left panel of
Figure~\ref{fig:evMpSh} shows the dependence of $\delmeb$ on the halo mass
of the primary cluster $M_\mathrm{pri}$ of the interacting pair. We control
the halo mass of the non-interacting clusters to be the same at each fixed
$M_\mathrm{pri}$ bin, so that any potential discrepancy should be caused by
the presence of a massive neighbour. In the low redshift bin~(blue filled
squares), although $\delmeb$ increases monotonically with $M_\mathrm{pri}$,
it is consistent with zero except for in the highest halo mass bin $\log
M_\mathrm{pri}{>}14.55$ where only nine interacting clusters are found. For
galaxies in the high redshift bin~(red filled triangles), the monotonic
trend disappears and we do not detect any evidence of positive $\delmeb$
across the entire halo mass range.  As a result, the overall $\delmeb$ from
combining the two redshift bins~(gray shaded band) is largely consistent
with zero at any fixed $M_\mathrm{pri}$, implying little to no impact of
cluster-scale tidal field on the bar strength.  Similarly, the right panel
of Figure~\ref{fig:evMpSh} shows the dependence of $\delmeb$ on the
projected cluster-centric distance scaled by the halo radius, $R/R_{200c}$.
Galaxies in the low redshift bin~(blue filled squares) exhibit a strong
enhancement in $\meb$ surrounding the interacting clusters at
$0.25{<}R/R_{200c}{<}0.5$, i.e., well within the virialised region of clusters.
However, this tantalizing signal of a positive $\delmeb$ disappears beyond
$R=R_{200c}$~(i.e., in the infall region outside the cluster radius) at
$0.015{<}z{<}0.06$ and across the entire range of $R/R_{200c}$ at
$0.06{<}z{<}0.105$~(red filled triangles).  Consequently, the overall signal
for galaxies at $0.015{<}z{<}0.105$~(gray shaded band) is consistent with zero
between the cluster centre and the infall region, exhibiting no signal of
tidal enhancement of bars.

\subsubsection{On the Discrepancy between the Observations at Low and High Redshifts}
\label{subsubsec:lowhigh}

In \S\ref{subsubsec:boost}, at $z<0.06$ we observe the enhanced average bar
strength in the vicinity of interacting clusters compared to the isolated
systems, as was firstly detected by \citetalias{Yoon2019}, but no such
signal for the larger sample at $z\ge 0.06$. This difference in the tidal
behaviour of bars at the low and high redshifts is intriguing, as it
requires a strong evolution in the formation timescale of tidally-induced
bars between $z\sim 0.1$ and today~(i.e., $\sim 1\,\mathrm{Gyr}$).  Since
the disc fraction does not vary since $z\sim0.1$~\citep{vanderKruit2011},
for such a strong evolution in tidal bars to occur, the average bar
formation timescale has to decrease dramatically to well below
$1\,\mathrm{Gyr}$.  Simulations generally predict that the formation
timescale of bars depends exponentially on the disc-to-total mass ratio
$f_{\mathrm{disc}}$~\citep{Fujii2018}, reaching below $1\,\mathrm{Gyr}$ for
galaxies with $f_{\mathrm{disc}}>0.5$. However, a rapid onset of bars due
to a boosted $f_{\mathrm{disc}}$ in strong tidal fields is highly unlikely
below $z\sim 0.1$, but only plausible in the very high-redshift
Universe~\citep{BlandHawthorn2023}.

Therefore, the discrepancy between the two redshift bins is more likely
caused by observational uncertainties. For instance, the efficacy of our
bar detection method could diminish rapidly with redshift, thereby smearing
the signal that otherwise exists in the high redshift bin. This explanation
is plausible, but we have conducted comprehensive tests on the sensitivity
of our bar detection method to redshift in \S\ref{sec:det}, which
demonstrate that we can distinguish between the barred vs. unbarred
galaxies reasonably well up to $z{=}0.11$. In addition, the comparison
between galaxies in the interacting vs. non-interacting clusters is done at
fixed redshifts, so that any systematic trend of $\eb$ with redshift would
not affect our results.  Therefore, our analysis should be able to pick up
at least some of the signal within $0.06{<}z{<}0.11$ should it be as strong as
was detected at $z{<}0.06$, especially given the significant reduction in the
statistical uncertainties due to the much larger cluster sample.

Alternatively, the discrepant results could be simply due to a statistic
fluke in the local Universe below $z{=}0.06$, where the observed enhancement
of bar strength is primarily contributed by the five cluster pairs~(nine
clusters in total). In the follow-up paper, we will apply our bar detection
method to the DESI images of those galaxies, and investigate if the
discrepancy remains~(hence more likely a fluke) or could be resolved by the
deeper imaging data.

To briefly summarise our results in the current section, we confirm the
findings from \citetalias{Yoon2019} and detect an enhancement of the bar
strength surrounding interacting clusters at $0.015{<}z{<}0.06$ using our
automated bar detection method and the \citetalias{Yang2007} cluster
sample. Furthermore, we find that the enhancement primarily originates from
the boosted fraction of barred galaxies within the central region~($R{<}0.5
R_{200c}$) of the most massive cluster-cluster pairs~($\log
M_{200c}{>}14.55$). However, the enhancement seen in the local clusters below
$z{=}0.06$ disappears as more clusters~(and their associated galaxies) from
the higher redshift up to $z{=}0.11$ are included in our final analysis.

\subsection{Barred galaxies in the large-scale tidal environment}
\label{subsec:large_scale}

We now explore the possible connection between bar strength and tidal
environment on the larger scales, where linear theory predicts a
correlation between the tidal anisotropy and halo spin, a potential
facilitator of bar growth. For the analysis in this section, we switch to
the stellar mass-limited~($\log M_{*}>10$) galaxy sample that is
volume-complete within $z=[0.01,0.074]$ in \S\ref{sec:data} and estimate
the bar strength of each disc-dominated galaxy~($0{<}\BT{<}0.5$) using our
automated bar detection method.  Following \citet{Alam2019}, we measure the
spherical galaxy overdensity $\degt$ as well as the tidal anisotropy
parameter $\afiv$ surrounding each disc-dominated galaxy from the 3D
distribution of all galaxies in this sample, as described in
\S\ref{subsec:tidalenv}. In addition, \citet{Alam2019} showed that there is
no correlation between $\degt$ and $\afiv$~(see their Fig. 1), which we
have verified explicitly using our sample.

\begin{figure}
	\includegraphics[width=0.96\columnwidth]{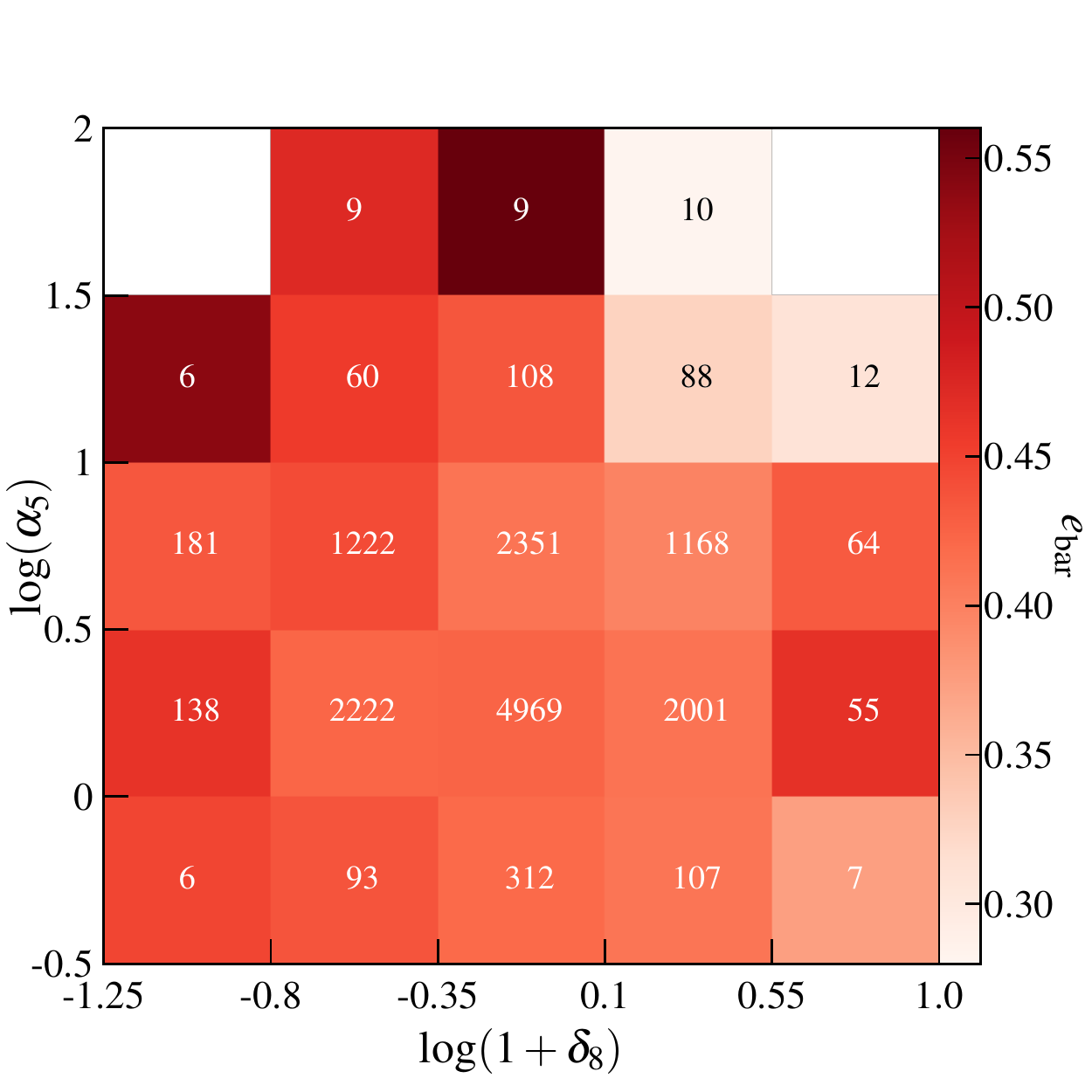}
    \caption{The average bar strength $\eb$ of galaxies on the
    $\log(1+\degt)$ vs.  $\log\afiv$ plane, colour-coded by the colour bar
    on the right. The number in each cell indicates the number of galaxies
    within that 2D bin.} \label{fig:d8_a5_hist2d}
\end{figure}

Figure~\ref{fig:skyscatter} provides a visually-appealing overview of the
different types of environments defined by $\degt$ or $\afiv$, showing the
spatial distribution of disc-dominated galaxies with $\eb \ge 0.375$~(red)
and $\eb < 0.375$~(blue) in the top/bottom $30\%$ of $\degt$~(left two
panels) and $\afiv$~(right two panels). To avoid clutter, we only select
galaxies from a thin redshift slice of $0.0535{<}z{<}0.0605$.  As expected,
galaxies in the high- and low-$\degt$ environments are densely and loosely
clustered, respectively, displaying drastically different levels of
clumpiness. Meanwhile, galaxy distributions in the high- and low-$\afiv$
environments have the similar clumpiness, but exhibit markedly different
levels of anisotropies on scales larger than ${\sim}10\,\hmpc$. Visually,
there is no discernible evidence of segregation between the more-barred vs.
less-barred galaxies in any of the four types of environments --- any tidal
dependence of bars, if exists, must be subtle and thus requires a more
quantitative investigation.

\begin{figure}
	\includegraphics[width=0.96\columnwidth]{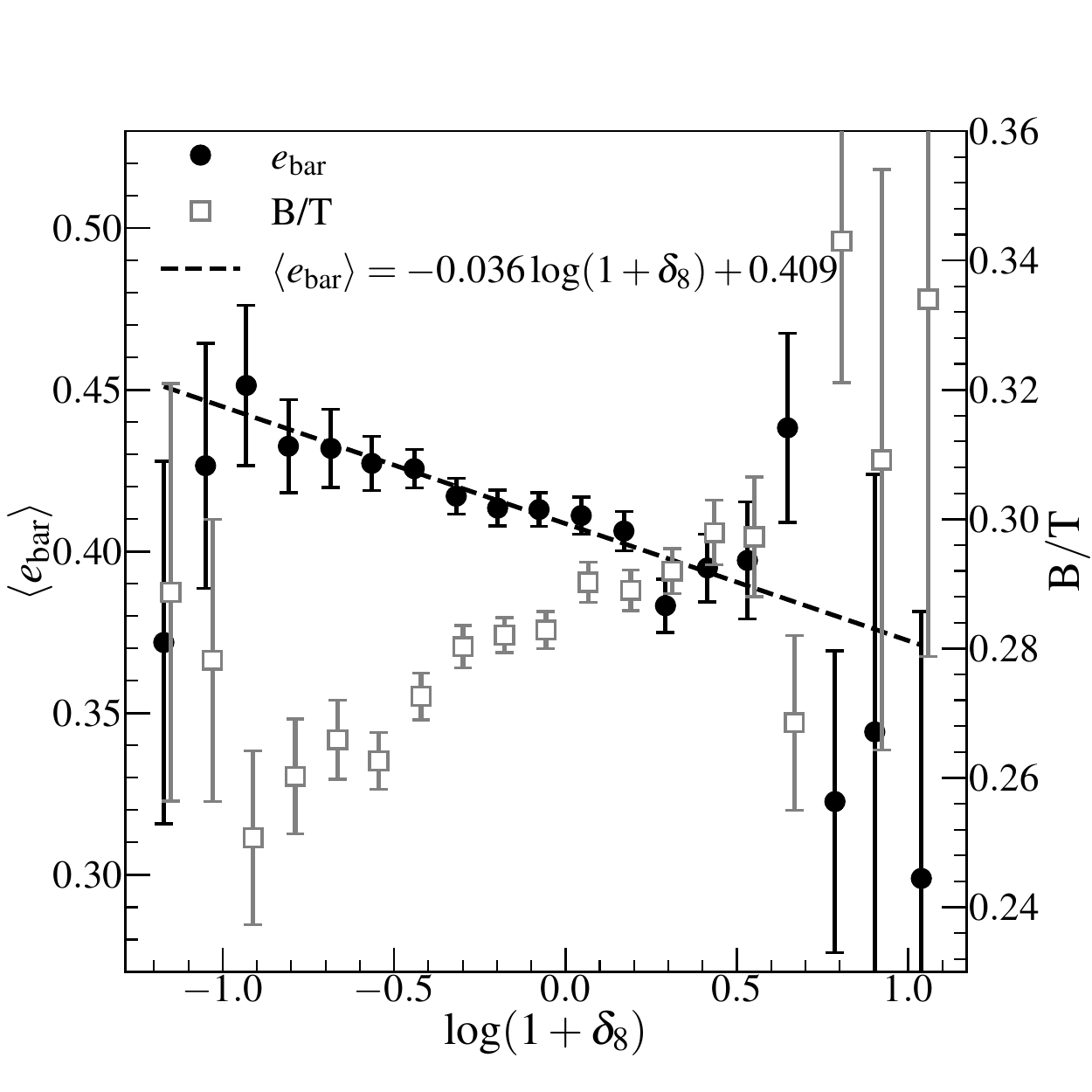}
    \caption{Dependence of mean bar strength $\langle\eb\rangle$~(black
    solid circles; left y-axis) and mean $\BT$~(gray open squares; right
    y-axis) on the spherical overdensity $\degt$. The black solid line is
    the best-fit linear relation between $\meb$ and $\log(1+\degt)$. All
    errorbars are $1{-}\sigma$ uncertainties estimated from Jackknife
    resampling.}
	\label{fig:evd8}
\end{figure}

We start our investigation with Figure~\ref{fig:d8_a5_hist2d}, which shows
the average bar strength $\meb$ of disc-dominated galaxies as a 2D function
of $\log(1+\degt)$ and $\log\afiv$, $\langle \eb | \afiv, \degt \rangle$,
colour-coded by the colourbar on the right. The number listed within each
cell indicates the number of galaxies within that 2D bin of fixed $\degt$
and $\afiv$. Overall, we observe a weak trend of declining $\meb$ with
$\log(1+\degt)$ in the horizontal direction, but a complex trend with
$\log\afiv$ in the vertical direction, especially at $\log\afiv>1$.  The
declining trend of $\meb$ with $\degt$ is presented more clearly in
Figure~\ref{fig:evd8}, where the filled circles~(open squares) with
errorbars show the dependence of $\meb$~($\BT$) of galaxies on
$\log(1+\degt)$ using the left~(right) y-axis. The declining trend of
$\meb$ with $\log(1+\degt)$ can be described by a simple linear relation as
\begin{equation}
    \langle \eb | \degt\rangle =-0.036 \log(1+\degt) + 0.409,
    \label{eqn:ebdelta}
\end{equation}
which is indicated by the black dashed line in Figure~\ref{fig:evd8}.  This
declining trend is likely caused by the strong anti-correlation between
$\eb$ and $\BT$, and since the $\BT$ of galaxies is higher in denser
environments~(open squares), it is unsurprising that the average bar
strength decreases with increasing $\degt$. This is consistent with the
previous studies that found no significant difference in terms of the local
density environment between barred and unbarred galaxies when other
physical properties of galaxies~(e.g., $\BT$) are
controlled~\citep{Aguerri2009,Li2009,Lee2012}.

\begin{figure}
	\includegraphics[width=0.96\columnwidth]{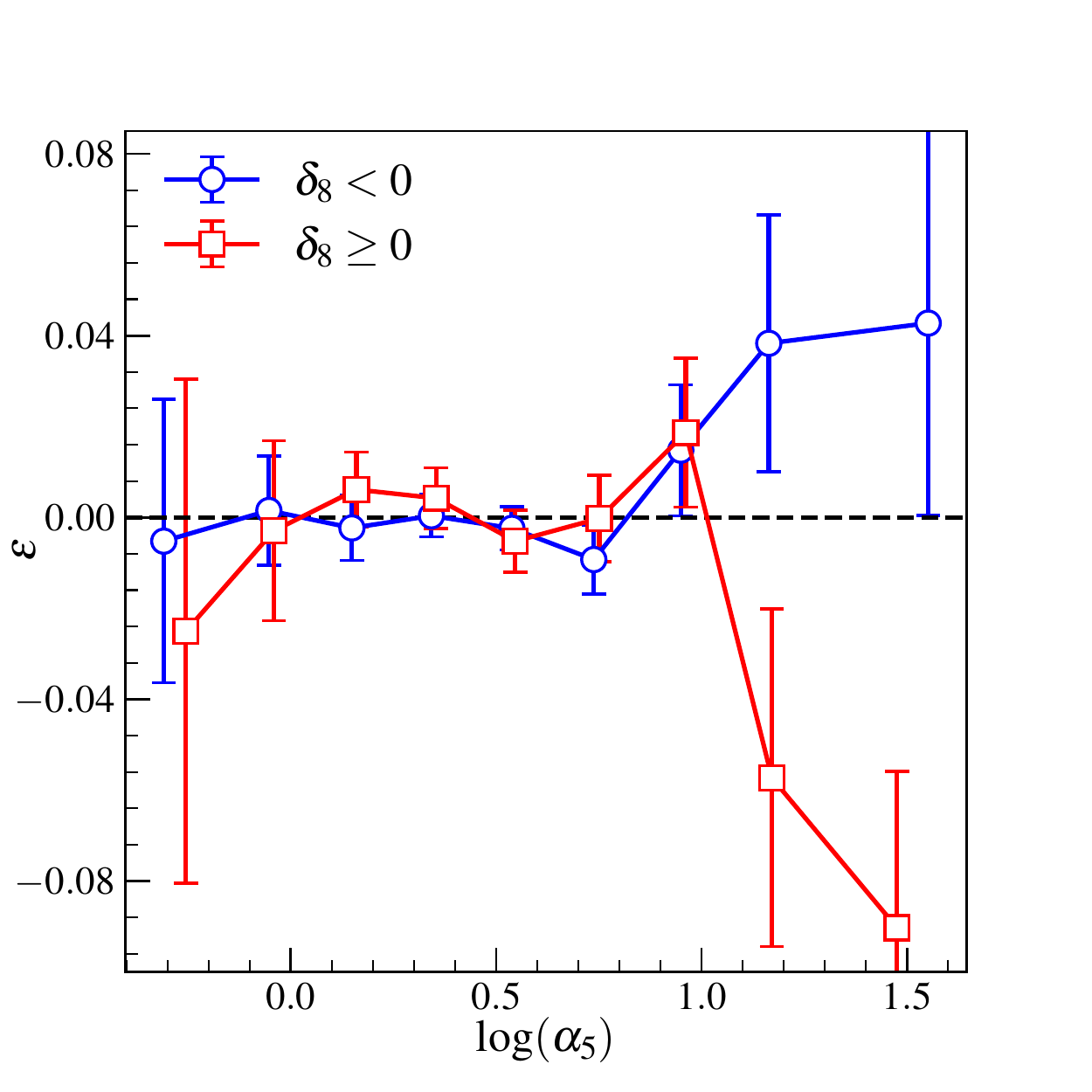}
    \caption{Dependence of the average excess bar strength
    $\varepsilon$~(defined in Equation~\ref{eqn:epsilon}) on $\afiv$ for
    galaxies in underdense~($\degt{<}0$; blue circles) and overdense~($\degt{\ge}
    0$; red squares) regions. All errorbars are $1-\sigma$
    uncertainties estimated from Jackknife resampling.}
	\label{fig:epva5}
\end{figure}

In order to disentangle the effect of tidal anisotropy on bar strength from
that of spherical overdensity, we remove the $\degt$-dependence of $\meb$
by defining the average excess bar strength $\varepsilon$ at fixed $\afiv$
and $\degt$ as \begin{equation} \varepsilon(\afiv, \degt) \equiv \langle
e_\mathrm{bar} \mid \afiv,\, \degt \rangle - \langle
e_\mathrm{bar}|\delta_{8} \rangle, \label{eqn:epsilon} \end{equation} where
$\langle \eb | \afiv, \degt \rangle$ is shown in
Figure~\ref{fig:d8_a5_hist2d} and $\langle e_\mathrm{bar}|\delta_{8}
\rangle$ is calculated from the linear relation of
Equation~\ref{eqn:ebdelta}. Figure~\ref{fig:epva5} shows the average excess
bar strength in the low- and high-$\degt$ environments, $\varepsilon(\afiv,
\degt < 0)$~(blue circles) and $\varepsilon(\afiv, \degt\ge 0)$~(red
squares), respectively, each as a function of $\log\afiv$.  The errorbars
are errors on the mean estimated from Jackknife resampling.  For the
majority~($98\%$) of galaxies that live in tidal environments with
$\log\afiv<1$, their excess bar strength is consistent with zero,
indicating that the anisotropy of their underlying tidal field does not
play a role in the formation of bars. Interestingly, in the extreme tidal
environments with $\log\afiv\ge 1$, the tidal dependence of bars in the
underdense~($\degt<0$) vs. overdense~($\degt\ge 0$) environments deviate
from zero and diverge in opposite directions --- the bar strength of
galaxies in the highly anisotropic, low-density regions is slightly
boosted, while the bar strength in the equally high-$\afiv$ but
high-density regions is somewhat hindered.

The straightforward interpretation is that the large-scale tidal anisotropy
has no impact on the formation and evolution of bars, except for the two
per cent of galaxies in the extreme tidal environments of $\log\afiv\ge1$.
In particular, fast-spinning haloes in those extreme tidal environments
promote bar formation in the underdense regions, consistent with the
predictions from simulations of galaxies in isolated
haloes~\citep{Saha2013}; Meanwhile in the overdense regions, fast-spinning
haloes may suppress the growth of bars, as predicted by the cosmological
simulations~\citep{Rosas-Guevara2022,Izquierdo-Villalba2022} where the
galaxies are embedded in the dense cosmic web.

However, although the non-zero signals at $\afiv{>}10$ is statistically
significant~(${>}1\sigma$), we caution that the Jackknife errorbars do not
include any potential systematic errors from cosmic variance or the tidal
anisotropy measurements. In particular, we use the same smoothing scale for
computing $\afiv$ in both the low- and high-density regions, which could
produce a differential aliasing effect that leads to some weak but
$\degt$-dependent correlation between $\afiv$ and $\degt$. Since such a
residual correlation cannot be removed by the estimator defined by
Equation~\ref{eqn:epsilon}, it could potentially mimic the signal observed
in Figure~\ref{fig:epva5}. Unfortunately, it is difficult to assess the
impact of such a potential systematic error with a relatively sparse survey
like the SDSS.  Therefore, although the large-scale tidal dependence of
bars is statistically detected above $1\sigma$ using the SDSS data, a
denser galaxy sample within a larger volume is required for a smoking-gun
detection of such an effect.

\section{Conclusions}
\label{sec:conc}

In this paper, we develop an automated bar detection method to measure the
bar strength $\eb$ of galaxies by applying the ellipse fitting over galaxy
images after subtracting the best-fitting 2D disc models.  Compared with
the conventional ellipse fitting scheme, our method is able to better
reveal the strength of the underlying bars that are otherwise embedded in
some of the bright discs. After performing an extensive suite of
comparisons with the visual identifications using either SDSS or DESI
imaging data, we find that our measurements of $\eb$ using disc-subtracted
images are robust against the decrease in image quality and spatial
resolution, and can thus be applied to SDSS images of galaxies up to
$z{=}0.11$, the maximum redshift of our analysis.

To investigate the dependence of $\eb$ on the small-scale tidal
environment, we make use of the cluster sample derived by the
\citetalias{Yang2007} group catalogue from SDSS. Following the recent study
of \citetalias{Yoon2019}, we divide the \citetalias{Yang2007} clusters into
interacting vs. non-interacting systems, and measure the difference between
the average bar strength $\langle\eb\rangle$ of galaxies surrounding
interacting clusters and that around isolated ones. Within the same
redshift range~($0.01{<}z{<}0.06$) probed by \citetalias{Yoon2019}, we
confirm their results that the $\langle\eb\rangle$ in interacting clusters
is higher than that in isolated systems. By examining the dependence of
such enhancement on cluster mass and projected distance to the cluster
centre, we find that the signal within $0.01{<}z{<}0.06$ is primarily
contributed by galaxies in the central regions~($R{<}0.5 R_{200c}$) of the
few very massive cluster-cluster pairs~($\log M_{200c}{>}14.55$). However,
after we increase the cluster sample by a factor of five by extending the
analysis up to $z{=}0.11$, the tidal enhancement of bars in the interacting
clusters goes away, indicating little correlation between the bar strength
and cluster-scale tidal strength. This small-scale tidal analysis can be
presumably extended to higher redshifts using photometric cluster
catalogues~\citep{Golden-Marx2022}, but the identification of cluster
member galaxies will be subjected to strong projection effects~\citep{Zu2017}.

For characterising the large-scale tidal environments, we adopt the tidal
anisotropy parameter $\afiv$ calculated from the overdensity field smoothed
over a scale of $5\,\hmpc$~\citep{Paranjape2018}. Assuming a large-scale
tidal origin~(at least partially) of the angular momentum of the haloes, we
use $\afiv$ as a proxy for the spin of haloes in different anisotropic
environments~\citep{Ramakrishnan2019}. Following \citet{Alam2019}, we
compute $\afiv$ from a stellar mass-limited~($\log M_*{>}10$) galaxy sample
between $0.01{<}z{<}0.074$, and measure the dependence of average bar
strength $\langle\eb\rangle$ on $\afiv$ at fixed spherical~(isotropic)
overdensities. We do not detect any such dependence for $98\%$ of the
galaxies residing in the environments with $\afiv{<}10$.
Intriguingly, among the $2\%$ with
$\alpha_{5}{\ge}10$, there is a hint of bar enhancement in the underdense
regions, where the disc-halo systems are more likely to evolve in
isolation. Since halo spin is correlated with the tidal anisotropy,
this bar enhancement is consistent with the prediction by \citet{Saha2013}
that halo spin promotes bar formation/growth. In contrast, galaxies in the
overdense regions exhibit suppressed bar strengths in the extremely
anisotropic environments with $\alpha_{5}{\ge}10$, consistent with the
prediction of some of the cosmological hydrodynamic
simulations~\citep{Rosas-Guevara2022,Izquierdo-Villalba2022}.

However, the non-zero signal at $\alpha_{5}{\ge}10$ is subjected to cosmic
variance and systematic uncertainties associated with the tidal anisotropy
measurements. For the cosmic variance, it could potentially be mitigated by
using a constrained simulation~\citep[e.g., ELUCID;][]{Wang2014} that
accurately reproduces the underlying density field within the SDSS
volume~\citep[see][for a similar application]{Salcedo2022}. Looking to the
future, both types of systematic errors can be better mitigated with the
Bright Galaxy Survey~\citep{Hahn2022} within the Dark Energy Spectroscopic
Instrument Survey~\citep[DESI;][]{DESI2022}. Meanwhile, our automated bar detection
method can be easily applied to upcoming space-based imaging surveys like
the Chinese Survey Space Telescope~\citep[{\it CSST};][]{gon19} and the
Roman Space Telescope~\citep[{\it Roman};][]{spe15}, both of which will
provide sharp images of galactic bars within a cosmological volume up to
much higher redshifts.

Combining our results on both small and large scales, we do not detect any
strong evidence for the dependence of bar strength on the tidal field.
Therefore, any tidal impact of bar formation, if exists, should be very
weak in the local Universe. Together with the general lack of bar
dependence on the overdensity environment~\cite{Aguerri2009, Lee2012,
Li2009, Skibba2012, Fraser-McKelvie2020}, our conclusion has important
implications for the theoretical understanding of bar formation --- the
primary driver of bar strength is most likely intrinsic to the disc galaxy
itself, rather than the tidal environment, whether it be interacting
clusters or tidal anisotropy.

\section*{Acknowledgements}

We thank Min Du, Sandeep Kumar Kataria, Zhaoyu Li, Zhi Li, and Juntai Shen for
helpful discussions. This work is supported by the National Key Basic
Research and Development Program of China (No. 2018YFA0404504), the
National Science Foundation of China (12173024, 11890692, 11873038,
11621303), the China Manned Space Project (No.  CMS-CSST-2021-A01,
CMS-CSST-2021-A02, CMS-CSST-2021-B01), and the ``111'' project of the
Ministry of Education under grant No. B20019. Y.Z. acknowledges the
generous sponsorship from Yangyang Development Fund. Y.Z. thanks Cathy
Huang for her hospitality during the pandemic and benifited greatly from
the stimulating discussions on bars at the Tsung-Dao Lee Institute.

\section*{Data Availability}

The data underlying this article will be shared on reasonable request to the corresponding author.



\bibliographystyle{mnras}
\bibliography{bar} 






\bsp	
\label{lastpage}
\end{document}